# Safe Privatization in Transactional Memory

Artem Khyzha
IMDEA Software Institute
Universidad Politécnica de Madrid

Alexey Gotsman
IMDEA Software Institute

Hagit Attiya
Technion

Noam Rinetzky
Tel-Aviv University

## Abstract

*Transactional memory* (TM) facilitates the development of concurrent applications by letting the programmer designate certain code blocks as atomic. Programmers using a TM often would like to access the same data both inside and outside transactions, e.g., to improve performance or to support legacy code. In this case, programmers would ideally like the TM to guarantee *strong atomicity*, where transactions can be viewed as executing atomically also with respect to non-transactional accesses. Since guaranteeing strong atomicity for arbitrary programs is prohibitively expensive, researchers have suggested guaranteeing it only for certain *data-race free (DRF)* programs, particularly those that follow the *privatization* idiom: from some point on, threads agree that a given object can be accessed non-transactionally. Supporting privatization safely in a TM is nontrivial, because this often requires correctly inserting *transactional fences*, which wait until all active transactions complete.

Unfortunately, there is currently no consensus on a single definition of transactional DRF, in particular, because no existing notion of DRF takes into account transactional fences. In this paper we propose such a notion and prove that, if a TM satisfies a certain condition generalizing opacity and a program using it is DRF *assuming* strong atomicity, then the program indeed has strongly atomic semantics. We show that our DRF notion allows the programmer to use privatization idioms. We also propose a method for proving our generalization of opacity and apply it to the TL2 TM.

## 1 Introduction

*Transactional memory* (TM) facilitates the development of concurrent applications by letting the programmer designate certain code blocks as *atomic* [22]. TM allows developing a program and reasoning about its correctness as if each atomic block executes as a *transaction*—atomically and without interleaving with other blocks—even though in reality the blocks can be executed concurrently. This guarantee can be formalized as *observational refinement* [7]: every behavior a user can observe of a program using a TM implementation can also be observed when the program uses an abstract TM that executes each block atomically. A TM can be implemented in hardware [23, 27], software [38] or a combination of both [11, 25].

Often programmers using a TM would like to access the same data both inside and outside transactions. This may be desirable to avoid performance overheads of transactional accesses, to support legacy code, or for explicit memory deallocation. One typical pattern is *privatization* [40], illustrated in Figure 1(a). There the atomic blocks return a value signifying whether the transaction committed or aborted. In the program, an object x is guarded by a flag x_is_private, showing whether the object should be accessed transactionally (false) or non-transactionally (true). The left-hand-side thread first tries to set the flag inside transaction $T_1$, thereby *privatizing* x. If successful, it then accesses x non-transactionally. A concurrent transaction $T_2$ in the right-hand-side thread checks the flag x_is_private prior to accessing x, to avoid simultaneous transactional and non-transactional access to the object. We expect the postcondition shown to hold: if privatization is successful, at the end of the program x should store 1, not 42. The opposite idiom is *publication*, illustrated in Figure 2. The left-hand-side thread writes to x non-transactionally and then clears the flag x_is_private inside transaction $T_1$, thereby *publishing*

(a) Delayed commit problem:

$$\{\, \mathtt{x\_is\_private} = \mathtt{false} \wedge \mathtt{x} = 0 \,\}$$

```
l := atomic { // T₁     atomic { // T₂
  x_is_private = true;    if (!x_is_private) {
}                           x = 42;
if (l == committed)       }
  x = 1; // v           }
```

$$\{\, l = \mathtt{committed} \implies \mathtt{x} = 1 \,\}$$

(b) Doomed transaction problem:

$$\{\, \mathtt{x\_is\_private} = \mathtt{false} \wedge \mathtt{x} = 0 \,\}$$

```
l := atomic { // T₁     atomic { // T₂
  x_is_private = true;    if (!x_is_private) {
}                           while (x == 1) {}
if (l == committed)       }
  x = 1; // v           }
```

**Figure 1.** Privatization examples.

$$\{\, \mathtt{x\_is\_private} = \mathtt{true} \wedge \mathtt{x} = l = 0 \,\}$$

```
x := 42; // v            l2 := atomic { // T₂
l1 := atomic { // T₁       if (!x_is_private)
  x_is_private = false;      l = x;
}                        }
```

$$\{\, l2 = \mathtt{committed} \wedge l \neq 0 \implies l = 42 \,\}$$

**Figure 2.** Publication example.

$$\{\, \mathtt{x} = \mathtt{y} = l1 = l2 = 0 \,\}$$

```
l := atomic { // T
  x := 1;                l1 := x; // v₁
  y := 2;                l2 := y; // v₂
}
```

$$\{\, \mathtt{x} = l1 \implies \mathtt{y} = l2 \,\}$$

**Figure 3.** A racy example.

x. The right-hand-side thread tests the flag inside transaction $T_2$, and if it is cleared, reads x. Again, we expect the postcondition to hold: if the right-hand-side thread sees the write to the flag, it should also see the write to x. The two idioms can be combined: the programmer may privatize an object, then access it non-transactionally, and then publish it back for transactional access.

Ideally, programmers mixing transactional and non-transactional accesses to objects would like the TM to guarantee *strong atomicity* [21], where transactions can be viewed as executing atomically also with respect to non-transactional accesses, i.e., without interleaving with them. This is equivalent to considering every non-transactional access as a single-instruction transaction. For example, the program in Figure 3 under strongly atomic semantics can only produce executions where each of the non-transactional accesses $v_1$ and $v_2$ executes either before or after the transaction $T$, so that the postcondition in Figure 3 always holds.

Unfortunately, providing strong atomicity in software requires instrumenting non-transactional accesses with additional instructions for maintaining TM metadata. This undermines scalability and makes it difficult to reuse legacy code. Since most existing TMs are either software-based or rely on a software fall-back, they do not perform such instrumentation and, hence, provide weaker atomicity guarantees. For example, they may allow the program in Figure 3 to execute non-transactional accesses $v_1$ and $v_2$ between transactional writes to x and y and, thus, observe an intermediate state of the transaction, e.g., x = 1 and y = 0, which violates the postcondition in Figure 3.

Researchers have suggested resolving the tension between strong TM semantics and performance by taking inspiration from non-transactional shared-memory models, which are subject to the same problem: optimizations performed by processors and compilers weaken the guarantee of *sequential consistency* [26] ideal for this setting. The compromise taken is to guarantee sequential consistency for certain *data-race free (DRF)* programs, which do not access the same data concurrently without synchronization [6]. Racy programs either are allowed to produce non-sequentially-consistent behaviors [29], or are declared faulty and thus having no semantics at all [2]. DRF thus establishes a contract between the programmer and the run-time system, which can be formalized by the so-called *Fundamental Property*: if a program is DRF *assuming* the strong semantics (such as sequential consistency), then the program does have the strong semantics. The crucial feature of this property is that DRF is checked by considering only executions under the strong semantics, which relieves the programmer from having to reason about the weaker semantics of unrestricted programs. The DRF contract has formed the basis of the memory models of both Java [29] and C/C++ [2].

Applying the above approach to TM, strong atomicity would be guaranteed only for programs that do not have an analog of data races in this setting—informally, concurrent transactional and non-transactional accesses to the same data [4, 5, 8, 9, 33, 40]. For example, we do not want to guarantee strong atomicity for the program in Figure 3, which has such concurrent accesses to x and y. On the other hand, the programs in Figure 1 and Figure 2 should be guaranteed strong atomicity, since at any point of time, an object is accessed either only transactionally or only non-transactionally. Unfortunately, whereas the DRF contract in non-transactional memory models has been worked out in detail, the situation in transactional models remains unsettled. There is currently no consensus on a single definition of transactional DRF: there are multiple competing proposals [4, 8, 9, 24, 28], which often come without a formal justification similar to the Fundamental Property of non-transactional memory models.

This paper makes a step towards a definition of transactional DRF on a par with solutions in non-transactional memory models. A key technical challenge we tackle is that many TM implementations, when used out-of-the-box, do not guarantee strong atomicity for seemingly well-behaved programs using privatization, such as the ones in Figure 1 [12, 15, 30, 37]. For example, such TMs may invalidate the postcondition of the program in Figure 1(a) due to the *delayed commit* problem [40]. In more detail, many software TM implementations execute transactions optimistically, buffering their writes, and flush them to memory only on commit. In this case, it is possible for the transaction $T_1$ to privatize x and for $v$ to modify it after $T_2$ started committing, but before its write to x reached the memory, so that $T_2$'s write subsequently overwrites $v$'s write and violates the postcondition. TMs that make transactional updates in-place and undo them on abort are subject to a similar problem. In Figure 1(b) we give another privatization example that is prone to a different problem—the *doomed transaction* problem [40]. A TM may execute $T_2$'s read from x_is_private, and then $T_1$ and $v$. Because $T_1$ modifies x_is_private, at this point $T_2$ is "doomed", i.e., guaranteed to abort if it finishes executing. But if the non-transactional write $v$ is uninstrumented and ignores the metadata the TM maintains to ensure the consistency of reads, $T_2$ will read the value written to x by $v$ and enter an infinite loop. This would never happen under strong atomicity, where $T_1$ and $v$ may not execute while $T_2$ is running.

A possible solution to the above problems is for the compiler or the programmer to insert special *transactional fences* [40]. These have semantics similar to *read-copy-update (RCU)* [31]: a fence blocks until all the transactions that were active when it was invoked complete, by either committing or aborting. For example, assume we insert a fence between the transaction $T_1$ and the non-transactional access $v$ in Figure 1(a). Then the delayed commit problem does not arise: if $T_2$ enters the if body and writes to x, then it must begin before the fence does; thus the fence will wait until $T_2$ completes and flushes its write to memory, so that $T_2$ cannot incorrectly overwrite $v$. Analogously, a fence between $T_1$ and $v$ in Figure 1(b) ensures that the doomed transaction problem does not arise: if $T_2$ reaches the while loop, then $v$ cannot execute before $T_2$ finishes, and thus the while loop immediately terminates.

Unfortunately, inserting transactional fences conservatively after every transaction, even when not required, undermines scalability. For example, Yoo et al. [42] showed that unnecessarily fencing a selection of transactional benchmarks leads to overheads of 32% on average and 107% in the worst case, the latter on one of the STAMP benchmarks [32]. For this reason, researchers have suggested placing transactional fences selectively, e.g., according to programmer annotations [42]. However, omitting fences without violating strong atomicity is nontrivial: for example, for several years the TM implementation in the GCC compiler had a buggy placement of transactional fences that omitted them after read-only transactions; this has recently been shown to violate strong atomicity [43]. To make sure such bugs are not habitual, we need a notion of transactional DRF that would take into account selective fence placements.

In this paper we propose just such a notion and formalize its Fundamental Property using observational refinement: if a program is DRF under strong atomicity (formalized as *transactional sequential consistency* [8, 9]), then all its executions are observationally equivalent to strongly atomic ones. We furthermore prove that the Fundamental Property holds under a certain condition on the TM, generalizing opacity [19, 20], which we call *strong opacity*. Thus, similarly to non-transactional memory models, the programmer writing code that has no data races according to our notion never needs to reason about weakly atomic semantics.

Our results thus establish a contract between client programs and TM implementations sufficient for strong atomicity. Of course, for this contract to be useful, it should not overconstrain either of its participants: programmers should be able to use the typical programming idioms, and common TM implementations should satisfy strong opacity that we require. In this paper we argue that this is indeed the case.

On the client side, our DRF notion allows the programmer to use privatization and publication idioms—programs in Figure 2 and in Figure 1 with a fence between $T_1$ and $v$ are considered data-race free and thus guaranteed strong atomicity. We hence view privatization and publication idioms as just particular ways of ensuring data-race freedom.

To justify appropriateness of our requirements on TM systems, we develop a method for proving that a TM satisfies strong opacity for DRF programs. Our method is *modular*: it requires only a minimal adjustment to a proof of the usual opacity of the TM assuming no mixed transactional/non-transactional accesses. We demonstrate the effectiveness of the method by applying it to prove the strong opacity of a realistic TM, TL2 [12], enhanced with transactional fences implemented using RCU. Our proof shows that this TM will indeed guarantee strong atomicity to programs satisfying our notion of DRF.

Thus, this paper makes the first proposal of transactional DRF that considers a flexible programming model (with transactional fences) and comes with a formal justification of its appropriateness—the Fundamental Property and the notion of TM correctness required for it to hold.

## 2 Programming Language

We now introduce a simple programming language with mixed transactional/non-transactional accesses, for which we formalize our results. We also define the semantics of the language when using a given TM implementation. As a special case of this semantics, we get the notion of strong atomicity [21] (also called transactional sequential consistency [8]): this is obtained by instantiating our semantics with a special idealized *atomic* TM where the execution of transactions does not interleave with that of other transactions or with non-transactional accesses.

### 2.1 Programming Language Syntax

A *program* $P = C_1 \parallel \ldots \parallel C_N$ in our language is a parallel composition of *commands* $C_t$ executed by different *threads* $t \in \text{ThreadID} = \{1, \ldots, N\}$. Every thread $t \in \text{ThreadID}$ has a set of *local variables* $l \in \text{LVar}_t$, which only it can access; for simplicity, we assume that these are integer-valued. Threads have access to a *transactional memory* (*TM*), which manages a fixed collection of *shared register objects* $x \in \text{Reg}$. The syntax of commands $C \in \text{Com}$ is as follows:

$$C = c \mid C \, ; C \mid \text{if } (b) \text{ then } C \text{ else } C \mid \text{while } (b) \text{ do } C$$
$$\mid l := \text{atomic}\{C\} \mid l := x.\text{read}() \mid x.\text{write}(e) \mid \text{fence}$$

where $b$ and $e$ denote Boolean, respectively, integer *expressions* over local variables and constants. The language includes *primitive commands* $c \in \text{PComm}$, which operate on local variables, and standard control-flow constructs.

An *atomic block* $l := \text{atomic}\{C\}$ executes $C$ as a *transaction*, which the TM can *commit* or *abort*. The system's decision is returned in the local variable $l$, which gets assigned a distinguished value committed or aborted. We do not allow programs to abort a transaction explicitly, and forbid nested atomic blocks and, hence, nested transactions.

Commands can invoke two methods on a shared register $x$: $x.\text{read}()$ returns the current value of $x$, and $x.\text{write}(e)$ sets it to $e$. Threads may call these methods both *inside* and *outside* atomic blocks. We refer to the former as a *transactional* accesses and to the latter as a *non-transactional* accesses. To make our presentation more approachable, following [20] we assume that each write in a single program execution writes a distinct value. Finally, the language includes a *transactional fence* command fence, which acts as explained in §1. It may only be used outside transactions.

The simplicity of the above language allows us to clearly explain our contributions. We leave handling advanced features, such as nested transactions [34, 35] and nested synchronization [39] as future work.

### 2.2 A Trace-based Model of Computations

To define the semantics of our programming language, we need a formal model for program computations. To this end, we introduce *traces*—certain finite sequences of *actions*, each describing a single computation step (for simplicity, in this paper we consider only finite computations). Let ActionId be a set of *action identifiers*. Actions are of two kinds. A *primitive action* denotes the execution of a primitive command and is of the form $(a, t, c)$, where $a \in \text{ActionId}$, $t \in \text{ThreadID}$ and $c \in \text{PComm}$. A *TM interface action* has one of the forms shown in Figure 4. We use $\alpha$ to range over actions.

| Request actions | Matching response actions |
|---|---|
| $(a, t, \text{txbegin})$ | $(a, t, \text{ok}) \mid (a, t, \text{aborted})$ |
| $(a, t, \text{txcommit})$ | $(a, t, \text{committed}) \mid (a, t, \text{aborted})$ |
| $(a, t, \text{write}(x, v))$ | $(a, t, \text{ret}(\bot)) \mid (a, t, \text{aborted})$ |
| $(a, t, \text{read}(x))$ | $(a, t, \text{ret}(v)) \mid (a, t, \text{aborted})$ |
| $(a, t, \text{fbegin})$ | $(a, t, \text{fend})$ |

**Figure 4.** TM interface actions. Here $a \in \text{ActionId}$, $t \in \text{ThreadID}$, $x \in \text{Reg}$, and $v \in \mathbb{Z}$.

TM interface actions denote the control flow of a thread $t$ crossing the boundary between the program and the TM: *request* actions correspond to the control being transferred from the former to the latter, and *response* actions, the other way around. A txbegin action is generated upon entering an atomic block, and a txcommit action when a transaction tries to commit upon exiting an atomic block. The request actions write$(x, v)$ and read$(x)$ denote invocations of the write, respectively, read methods of register $x$; a write action is annotated with the value $v$ written. The response actions ret$(\bot)$ and ret$(v)$ denote the return from invocations of write, respectively, read methods of a register; the latter is annotated with the value $v$ read. For reasons explained below, we consider non-transactional accesses to registers as calling into the TM, and hence use the same actions for them as for transactional accesses. The TM may abort a transaction (but not a non-transactional access) at any point when it is in control; this is recorded by an aborted response action. The actions fbegin and fend denote the beginning, respectively, the end of the execution of a fence command. In the following _ denotes an irrelevant expression.

**Definition 2.1.** A *trace* $\tau$ is a finite sequence of actions satisfying certain well-formedness conditions (stated informally due to space constraints; see §A.1):

- every action in $\tau$ has a unique identifier;
- commands in actions executed by a thread $t$ do not access local variables of other threads $t' \neq t$;
- every write operation writes a unique value distinct from $v_{\text{init}}$ (the initial value of each register);
- for every thread $t$, the projection $\tau|_t$ of $\tau$ onto the actions by $t$ cannot contain a request action immediately followed by a primitive action;
- request and response actions are properly matched;
- actions denoting the beginning and the end of transactions are properly matched;
- non-transactional accesses execute atomically: if $\tau = \tau_1 \, \alpha \, \tau_2$, where $\alpha$ is a read or a write request action by thread $t$, and all the transactions of $t$ in $\tau_1$ completed, then $\tau_2$ begins with a response to $\alpha$.
- non-transactional accesses never abort;
- fence actions may not occur inside transactions; and

- fence blocks until all active transactions complete: if $\tau = \tau_1 \,(\_, t, \text{txbegin})\, \tau_2 \,(\_, t', \text{fbegin})\, \tau_3 \,(\_, t', \text{fend})\, \tau_4$ then either $\tau_2$ or $\tau_3$ contains an action of the form $(\_, t, \text{committed})$ or $(\_, t, \text{aborted})$.

We denote the set of traces by Trace and the set of actions in a trace $\tau$ by $\text{act}(\tau)$. For a trace $\tau = \tau_0\_$, where $\tau_0$ is also a trace, we say that $\tau_0$ is a prefix of $\tau$.

A *transaction* $T$ is a nonempty trace such that it contains actions by the same thread, begins with a txbegin action and only its last action can be a committed or an aborted action. A transaction $T$ is: *committed* if it ends with a committed action, *aborted* if it ends with aborted, *commit-pending* if it ends with txcommit, and *live*, in all other cases. A transaction $T$ is in a trace $\tau$ if $T$ is a subsequence of $\tau$ and no longer transaction is. We let $\text{txns}(\tau)$ be the set of transactions in $\tau$.

We refer to TM interface actions in a trace outside of a transaction as *non-transactional actions*. We call a matching request/response pair of a read or a write a *non-transactional access*. We denote by $\text{nontxn}(\tau)$ the set of non-transactional accesses in $\tau$ and range over them by $\nu$.

A *history* is a trace containing only TM interface actions; we use $H, S$ to range over histories. Since histories fully capture the possible interactions between a TM and a client program, we often conflate the notion of a TM and the set of histories it produces. Hence, a *transactional memory* $\mathcal{H}$ is a set of histories that is prefix-closed and closed under renaming action identifiers. Note that histories include actions corresponding to non-transactional accesses, even though these may not be directly managed by the TM implementation. This is needed to account for changes to registers performed by such actions when defining the TM semantics: e.g., in the case when a register is privatized, modified non-transactionally and then published back for transactional access. Of course, a well-formed TM semantics should not impose restrictions on the placement of non-transactional actions, since these are under the control of the program.

### 2.3 Programming Language Semantics

The *semantics* of the programming language is the set of traces that computations of programs produce. Due to space constraints, we defer its formal definition to §A.2 and describe only its high-level structure. A *state* of a program $P = C_1 \parallel \ldots \parallel C_N$ records the values of all its variables: $s \in \text{State} = (\biguplus_{t=1}^{N} \text{LVar}_t) \to \mathbb{Z}$. The semantics of a program $P$ is given by the set of traces $[\![P, \mathcal{H}]\!](s) \subseteq \text{Trace}$ it produces when executed with a TM $\mathcal{H}$ from an initial state $s$. To define this set, we first define the set of traces $[\![P]\!](s) \subseteq \text{Trace}$ that a program can produce when executed from $s$ with the behavior of the TM unrestricted, i.e., considering all possible values the TM can return on reads and allowing transactions to commit or abort arbitrarily. This definition follows the intuitive semantics of our programming language. We then restrict $[\![P]\!](s)$ to the set of traces produced by $P$ when executed with $\mathcal{H}$ by selecting those traces that interact with the TM in a way consistent with $\mathcal{H}$: $[\![P, \mathcal{H}]\!](s) = \{\tau \mid \tau \in [\![P]\!](s) \wedge \text{history}(\tau) \in \mathcal{H}\}$, where $\text{history}(\cdot)$ projects to TM interface actions.

### 2.4 Strong Atomicity

We now define an idealized *atomic* TM $\mathcal{H}_{\text{atomic}}$ where the execution of transactions does not interleave with that of other transactions or with non-transactional accesses. By instantiating the semantics of §2.3 with this TM, we formalize the strongly atomic semantics [21] (transactional sequential consistency [8, 9]). $\mathcal{H}_{\text{atomic}}$ contains only histories that are *non-interleaved*, i.e., where actions by one transaction do not overlap with the actions of another transaction or of non-transactional accesses. Note that by definition actions pertaining to different non-transactional accesses cannot interleave. Note also that transactions in a non-interleaved history do not have to be complete. For example,

$H_0 = (\_, t_1, \text{txbegin})\,(\_, t_1, \text{ok})\,(\_, t_1, \text{write}(x, 1))\,(\_, t_1, \text{ret}(\bot))$
$(\_, t_1, \text{txcommit})\,(\_, t_2, \text{txbegin})\,(\_, t_2, \text{ok})(\_, t_2, \text{write}(x, 2))$
$(\_, t_3, \text{read}(x))\,(\_, t_3, \text{ret}(1))$

is non-interleaved. We have to allow such histories in $\mathcal{H}_{\text{atomic}}$, because they may be produced by programs in our language, e.g., due to a non-terminating loop inside an atomic block.

We define $\mathcal{H}_{\text{atomic}}$ in such a way that the changes made by a live or aborted transaction are invisible to other transactions. However, there is no such certainty in the treatment of a commit-pending transaction: the TM implementation might have already reached a point at which it is decided that the transaction will commit. Then the transaction is effectively committed, and its operations may affect other transactions [20]. To account for this, when defining $\mathcal{H}_{\text{atomic}}$ we consider every possible completion of each commit-pending transaction in a history to either committed or an aborted one. Formally, we say that a history $H^c$ is a *completion* of a non-interleaved history $H$ if: (i) $H^c$ is non-interleaved; (ii) $H^c$ is has no commit-pending transactions; (iii) $H$ is a subsequence of $H^c$; and (iv) any action in $H^c$ which is not in $H$ is either a committed or an aborted action. For example, we can obtain a completion of history $H_0$ above by inserting $(\_, t_1, \text{committed})$ after $(\_, t_1, \text{txcommit})$.

We define $\mathcal{H}_{\text{atomic}}$ as the set of all non-interleaved histories $H$ that have a completion $H^c$ where every response action of a $\text{read}(x)$ returns the value $v$ in the last preceding $\text{write}(x, v)$ action that is not located in an aborted or live transaction different from the one of the read; if there is no such write, the read should return the initial value $v_{\text{init}}$. For example, $H_0 \in \mathcal{H}_{\text{atomic}}$. Hence, $\mathcal{H}_{\text{atomic}}$ defines the intuitive atomic semantics of transactions.

## 3 Data-Race Freedom

A data race happens between a pair of *conflicting* actions, as defined below.

**Definition 3.1.** A non-transactional request action $\alpha$ and a transactional request action $\alpha'$ *conflict* if $\alpha$ and $\alpha'$ are executed by different threads and they are read or write actions on the same register, with at least one being a write.

For such actions to form a data race, they should be *concurrent*. As is standard, we formalize this using a *happens-before* relation on actions in a history: $\mathrm{hb}(H) \subseteq \mathrm{act}(H) \times \mathrm{act}(H)$. To streamline explanations, we first define DRF in terms of happens-before, and only after this define the latter. For a history $H$ and an index $i$, let $H(i)$ denote the $i$-th action in the sequence $H$.

**Definition 3.2.** Actions $H(i)$ and $H(j)$ in a history $H$ form a *data race*, if they conflict and are not related by $\mathrm{hb}(H)$ either way. A history $H$ is *data-race free (DRF)*, written $\mathrm{DRF}(H)$, if it has no data races.

**Definition 3.3.** A program $P$ is *data-race free (DRF)* when executed from a state $s$ with a TM $\mathcal{H}$, written $\mathrm{DRF}(P, s, \mathcal{H})$, if $\forall \tau \in [\![P]\!](\mathcal{H}, s).\ \mathrm{DRF}(\mathrm{history}(\tau))$.

Our goal is to enable programmers to ensure strong atomicity of a program by checking its data-race freedom. However, the notion of DRF depends on the TM $\mathcal{H}$, and we do not want the programmer to have to reason about the actual TM implementation. In Section 5, we give conditions on TM $\mathcal{H}$ under which strong atomicity of a program is guaranteed if it is DRF *assuming* strong atomicity, i.e., $\mathrm{DRF}(P, s, \mathcal{H}_{\mathrm{atomic}})$ for $\mathcal{H}_{\mathrm{atomic}}$ from §2.4. We next define $\mathrm{hb}(H)$ and show examples of programs that are racy and race-free under $\mathcal{H}_{\mathrm{atomic}}$.

For a history $H$, we define several relations over $\mathrm{act}(H)$, which we explain in the following:

- *execution order*: $\alpha <_H \alpha'$ iff
  for some $i, j$ we have $\alpha = H(i), \alpha' = H(j)$ and $i < j$.
- *per-thread order* $\mathrm{po}(H)$: $\alpha <_{\mathrm{po}(H)} \alpha'$ iff
  $\alpha <_H \alpha'$ and actions $\alpha$ and $\alpha'$ are by the same thread.
- *restricted per-thread order* $\mathrm{xpo}(H)$: $\alpha <_{\mathrm{xpo}(H)} \alpha'$ iff
  $\alpha <_H \alpha'$, actions $\alpha$ and $\alpha'$ are by the same thread $t$, and there is a $(\_, t, \mathrm{txbegin})$ action between $\alpha$ and $\alpha'$.
- *client order* $\mathrm{cl}(H)$: $\alpha <_{\mathrm{cl}(H)} \alpha'$ iff
  $\alpha <_H \alpha'$ and $\alpha, \alpha'$ are non-transactional in $H$.
- *after-fence order* $\mathrm{af}(H)$: $\alpha <_{\mathrm{af}(H)} \alpha'$ iff
  $\alpha <_H \alpha', \alpha = (\_, \_, \mathrm{fbegin})$ and $\alpha' = (\_, \_, \mathrm{txbegin})$, i.e., the transaction begins after the fence does (Figure 5(a)).
- *before-fence order* $\mathrm{bf}(H)$: $\alpha <_{\mathrm{bf}(H)} \alpha'$ iff
  $\alpha <_H \alpha', \alpha \in \{(\_, \_, \mathrm{committed}), (\_, \_, \mathrm{aborted})\}$ and $\alpha' = (\_, \_, \mathrm{fend})$, i.e., the transaction ends before the fence does (Figure 5(b)).
- *read-dependency* relation $\mathrm{wr}_x(H)$ for $x \in \mathrm{Reg}$[1]:
  $\alpha <_{\mathrm{wr}_x(H)} \alpha'$ iff $\alpha = (\_, \_, \mathrm{write}(x, v)), \alpha' = (\_, \_, \mathrm{ret}(v))$ and the matching request action for $\alpha'$ is $(\_, \_, \mathrm{read}(x))$.

---

[1] The notation wr, standing for "write-to-read", is chosen to mirror other kinds of dependencies introduced in §6.

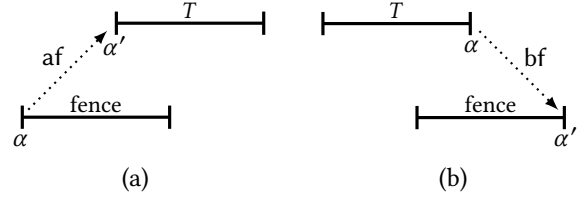

**Figure 5.** Illustration of the fence relations.

```
        { x_is_ready = false ∧ x = 0 }
l1 := atomic { // T       ║  do {
  x = 42;                 ║    l2 := x_is_ready; // v'
}                         ║  } while (¬l2);
x_is_ready := true; // v  ║  int l3 := x; // v''
        { l1 = committed  ⟹  l3 = 42 }
```

**Figure 6.** Privatization by agreement outside transactions.

- *transactional read-dependency* relation $\mathrm{txwr}_x(H)$:
  $\alpha <_{\mathrm{txwr}_x(H)} \alpha'$ iff $\alpha <_{\mathrm{wr}_x(H)} \alpha'$, and $\alpha$ and $\alpha'$ are transactional.

**Definition 3.4.** For a history $H$ we let the *happens-before* relation of $H$ be

$$\mathrm{hb}(H) = (\mathrm{po}(H) \cup \mathrm{cl}(H) \cup \mathrm{af}(H) \cup \mathrm{bf}(H) \cup$$
$$\bigcup_{x \in \mathrm{Reg}} (\mathrm{xpo}(H)\ ;\ \mathrm{txwr}_x(H)))^+.$$

Components of the happens-before describe various forms of synchronization available in our programming language, which we now explain one by one. First, actions by the same thread cannot be concurrent, and thus, we let $\mathrm{po}(H) \subseteq \mathrm{hb}(H)$. To concentrate on issues related to TM, in this paper we do not consider the integration of transactions into a language with a weak memory model and assume that the underlying non-transactional memory is sequentially consistent. Hence, we do not consider pairs of concurrent non-transactional accesses as races and let $\mathrm{cl}(H) \subseteq \mathrm{hb}(H)$. This can be used to privatize an object by agreeing on its status outside transactions, as illustrated in Figure 6. There the left-hand-side thread writes to x inside a transaction and then sets the flag x_is_ready outside. The right-hand-side thread keeps reading the flag non-transactionally until it is set, and then reads x non-transactionally. This program is DRF under $\mathcal{H}_{\mathrm{atomic}}$ because, in any of its traces, the conflicting write in $T$ and the non-transactional read $v''$ are ordered in happens-before due to the client order between the write in $v$ and the read in $v'$ that causes the do loop to terminate.

We also have $\mathrm{xpo}(H)\ ;\ \mathrm{txwr}_x(H) \subseteq \mathrm{hb}(H)$. Intuitively, this is because, if we have $(\alpha, \alpha') \in \mathrm{txwr}_x(H)$, then the commands by the thread of $\alpha$ preceding the transaction of $\alpha$ are guaranteed to have taken effect by the time $\alpha'$ executes.[2]

---

[2] Note, however, that the commands preceding $\alpha$ in its transaction may not have taken effect by this time: the TM may flush the writes of the

This ensures that publication can be done safely, as we now illustrate by showing that the program in Figure 2 is DRF under $\mathcal{H}_{\text{atomic}}$. Traces of the program may have only a single pair of conflicting actions—the accesses to x in $v$ and $T_2$. For both conflicting actions to occur, $T_2$ has to read false from x_is_private. Since under $\mathcal{H}_{\text{atomic}}$ transactions do not interleave with other transactions or non-transactional accesses, for this $T_1$ has to execute before $T_2$, yielding a history of the form $v\,T_1\,T_2$. In this history, we have a read-dependency between the write to x_is_private in $T_1$ and the read from x_is_private in $T_2$. But then the write to x in $v$ happens before the read from x in $T_2$, so that these actions cannot form a race.

Relations af($H$) and bf($H$) are used to formalize synchronization ensured by transactional fences. Recall that a fence blocks until all active transactions complete, by either committing or aborting. Hence, every transaction either begins after the fence does (and thus the fence does not need to wait for it; Figure 5(a)) or ends (including any required clean-up) before the fence does (Figure 5(b)). The relations af($H$) and bf($H$) capture the two respective cases. Note that, as required by Definition 2.1, every transaction has to be related to a fence at least by one of the two relations: a transaction may not span a fence.

Including after-fence and before-fence relations into happens-before ensures that privatization can be done safely given an appropriate placement of fences. To illustrate this, we show that the programs in Figure 1 are DRF under $\mathcal{H}_{\text{atomic}}$ when we place a transactional fence between $T_1$ and $v$. The possible conflicts are between the accesses to x in $v$ and $T_2$. For a conflict to occur, $T_2$ has to read false from x_is_private. Then $T_2$ has to execute before $T_1$, yielding a history $H$ of the form $T_2\,T_1\,\alpha_1\,\alpha_2\,v$, where $\alpha_1$ and $\alpha_2$ denote the request and the response actions of the fence. Since $T_2$ occurs before $\alpha_2$ in the history, they are related by the before-fence relation. But then the accesses to x in $T_2$ happen-before the write in $v$ and, therefore, the conflicting actions do not form a race. Finally, the program in Figure 3 is racy, since its traces contain pairs of conflicting actions unordered in happens-before. Inserting fences into this program will not make it DRF.

Our notion of DRF under $\mathcal{H}_{\text{atomic}}$ establishes the condition that a program has to satisfy to be guaranteed strong atomicity. In the next section, we formulate the obligations of its TM counterpart in the DRF contract.

## 4 Strong Opacity

We state the requirements on a TM $\mathcal{H}$ by generalizing the notion of *opacity* [19, 20], yielding what we call *strong opacity*. As part of our definition, we require that a history $H$ of a TM $\mathcal{H}$ can be matched by a history $S$ of the atomic TM $\mathcal{H}_{\text{atomic}}$

---

transaction to the memory in any order. This is why we do not require $\text{txwr}_x(H) \subseteq \text{hb}(H)$.

that "looks similar" to $H$ from the perspective of the program. The similarity is formalized by the following relation $H \sqsubseteq S$, which requires $S$ to be a permutation of $H$ preserving the happens-before relation.

**Definition 4.1.** A history $H_1$ is in the *strong opacity relation* with a history $H_2$, written $H_1 \sqsubseteq H_2$, if there is a bijection $\theta : \{1, \ldots, |H_1|\} \to \{1, \ldots, |H_2|\}$ such that:
- $\forall i.\, H_1(i) = H_2(\theta(i))$, and
- $\forall i, j.\, i < j \land H_1(i) <_{\text{hb}(H_1)} H_2(j) \implies \theta(i) < \theta(j)$.

The original definition of opacity requires any history of a TM $\mathcal{H}$ to have a matching history of the atomic TM $\mathcal{H}_{\text{atomic}}$. However, such a requirement would be too strong for our setting: since the TM has no control over non-transactional actions of its clients, histories in $\mathcal{H}$ may be racy, and we do not want to require the TM to guarantee strong atomicity in such cases. Hence, our definition of strong opacity requires only DRF histories to have justifications in $\mathcal{H}_{\text{atomic}}$. Let $\mathcal{H}|_{\text{DRF}} = \{H \in \mathcal{H} \mid \text{DRF}(H)\}$.

**Definition 4.2.** A TM $\mathcal{H}$ is *strongly opaque*, written $\mathcal{H}|_{\text{DRF}} \sqsubseteq \mathcal{H}_{\text{atomic}}$, if

$$\forall H.\, H \in \mathcal{H}|_{\text{DRF}} \implies \exists S.\, S \in \mathcal{H}_{\text{atomic}} \land H \sqsubseteq S.$$

Apart from the restriction to DRF histories, strong opacity and the usual opacity differ in several other ways. First, unlike in the usual opacity, our histories include non-transactional actions, because these can affect the behavior of the TM (e.g., via the idiom of "privatize, modify non-transactionally, publish", §2.2). Second, instead of preserving happens-before $\text{hb}(H_1)$ in Definition 4.1, the usual opacity requires preserving the program order $\text{po}(H_1)$ and the following *real-time order* $\text{rt}(H_1)$ on actions: $\alpha <_{\text{rt}(H)} \alpha'$ iff $\alpha \in \{(\_,\_,\text{committed}), (\_,\_,\text{aborted})\}, \alpha' = (\_,\_,\text{txbegin})$ and $\alpha <_H \alpha'$. This orders non-overlapping transactions, with the duration of a transaction determined by the interval from its txbegin action to the corresponding committed or aborted action (or to the end of the history if there is none). As shown in [16], preserving real-time order is unnecessary if program threads do not have means of communication not reflected in histories. Since we record the actions using both transactional and non-transactional accesses, preserving real-time order is unnecessary for our results. However, we use this order to prove strong opacity by adjusting the proofs of the usual one (§6). Finally, preserving happens-before in Definition 4.1 is required so that we could check DRF assuming strong atomicity, as we explain next.

## 5 The Fundamental Property

We now formalize the Fundamental Property of our DRF notion using *observational refinement* [7]: if a program is DRF under the atomic TM $\mathcal{H}_{\text{atomic}}$, then any trace of the program under a strongly opaque TM $\mathcal{H}$ has an *observationally equivalent* trace under the atomic TM $\mathcal{H}_{\text{atomic}}$.

**Definition 5.1.** Traces $\tau$ and $\tau'$ are *observationally equivalent*, denoted by $\tau \sim \tau'$, if

$$(\forall t \in \mathsf{ThreadID}.\ \tau|_t = \tau'|_t) \wedge (\tau|_{\mathsf{nontx}} = \tau'|_{\mathsf{nontx}}),$$

where $\tau|_{\mathsf{nontx}}$ denotes the subsequence of $\tau$ containing all actions from non-transactional accesses.

Equivalent traces are considered indistinguishable to the user. In particular, the sequences of non-transactional accesses in equivalent traces (which usually include all input-output) satisfy the same linear-time temporal properties. We lift the equivalence to sets of traces as follows.

**Definition 5.2.** A set of traces $\mathcal{T}$ *observationally refines* a set of traces $\mathcal{T}'$, written $\mathcal{T} \preceq \mathcal{T}'$, if $\forall \tau \in \mathcal{T}.\ \exists \tau' \in \mathcal{T}'.\ \tau \sim \tau'$.

**Theorem 5.3** (Fundamental Property). *If $\mathcal{H}$ is a TM such that $\mathcal{H}|_{\mathsf{DRF}} \sqsubseteq \mathcal{H}_{\mathsf{atomic}}$, then*

$$\forall P, s.\ \mathsf{DRF}(P, s, \mathcal{H}_{\mathsf{atomic}}) \implies [\![P]\!](\mathcal{H}, s) \preceq [\![P]\!](\mathcal{H}_{\mathsf{atomic}}, s).$$

Theorem 5.3 establishes a contract between the programmer and the TM implementors. The TM implementor has to ensure strong opacity of the TM assuming the program is DRF: $\mathcal{H}|_{\mathsf{DRF}} \sqsubseteq \mathcal{H}_{\mathsf{atomic}}$. The programmer has to ensure the DRF of the program assuming strong atomicity: $\mathsf{DRF}(P, s, \mathcal{H}_{\mathsf{atomic}})$. This contract lets the programmer to check properties of a program assuming strong atomicity ($[\![P]\!](\mathcal{H}_{\mathsf{atomic}}, s)$) and get the guarantee that the properties hold when the program uses the actual TM implementation ($[\![P]\!](\mathcal{H}, s)$). We have already shown that the expected privatization and publication idioms are DRF under strong atomicity (§3), so that the programmer can satisfy its part of the contract. In the following sections we develop a method for discharging the obligations of the TM.

The proof of Theorem 5.3 follows directly from the following lemma, proved in §B.1.

**Lemma 5.4.** *If $\mathcal{H}$ is a TM such that $\mathcal{H}|_{\mathsf{DRF}} \sqsubseteq \mathcal{H}_{\mathsf{atomic}}$, then:*
1. $\forall P, s.\ \mathsf{DRF}(P, s, \mathcal{H}) \implies [\![P]\!](\mathcal{H}, s) \preceq [\![P]\!](\mathcal{H}_{\mathsf{atomic}}, s).$
2. $\forall P, s.\ \mathsf{DRF}(P, s, \mathcal{H}_{\mathsf{atomic}}) \implies \mathsf{DRF}(P, s, \mathcal{H}).$

Part 1 shows that if a program is DRF under the concrete TM $\mathcal{H}$, then it has the expected strongly atomic semantics. It is an adaptation of a result from [7]. Part 2 enables checking DRF using an atomic TM $\mathcal{H}_{\mathsf{atomic}}$ and is a contribution of the present paper. Its proof relies on the fact that strong opacity preserves happens-before (Definition 4.1).

## 6 Proving Strong Opacity

We now develop a method for proving $\mathcal{H}|_{\mathsf{DRF}} \sqsubseteq \mathcal{H}_{\mathsf{atomic}}$. The method builds on a *graph characterization* of opacity of Guerraoui and Kapalka [20], which was proposed for proving the usual opacity of TMs that do not allow mixed transactional/non-transactional accesses to the same data. The characterization allows checking opacity of a history by checking two properties: *consistency* of the history and the acyclicity of a certain *opacity graph*, which we define further in this section.

Consistency captures some very basic properties of read-dependency relation $\mathsf{wr}_x$ (for each register $x$) that have to be satisfied by every opaque TM history. Intuitively, in a consistent history every transaction $T$ reading the value of a register $x$ either reads the latest value $T$ itself wrote to $x$ before, or some value written non-transactionally or by a committed or commit-pending transaction.

**Definition 6.1.** A pair of matching request and response actions $(\alpha, \alpha')$ is said to be *local* to $T \in \mathsf{txns}(H)$, if:
- $\alpha = (\_, \_, \mathsf{read}(x)) \wedge$
  $\exists \beta \in T.\ \beta <_{\mathsf{po}(H)} \alpha \wedge \beta = (\_, \_, \mathsf{write}(x, \_))$; or
- $\alpha = (\_, \_, \mathsf{write}(x, \_)) \wedge$
  $\exists \beta \in T.\ \alpha <_{\mathsf{po}(H)} \beta \wedge \beta = (\_, \_, \mathsf{write}(x, \_))$.

We let $\mathsf{local}(H)$ denote the set of all local actions in $H$.

Thus, local reads from $x$ are preceded by a write to $x$ in the same transaction; local writes to $x$ are followed by a write to $x$ in the same transaction.

**Definition 6.2.** In a history $H$, a read request $\alpha = (\_, \_, \mathsf{read}(x))$ and its matching response $\alpha' = (\_, \_, \mathsf{ret}(v))$ are said to be *consistent*, if:
- when $(\alpha, \alpha') \in \mathsf{local}(H)$ and performed by a transaction $T$, $v$ is the value written by the most recent write $(\_, \_, \mathsf{write}(x, v))$ preceding the read in $T$;
- when $(\alpha, \alpha') \notin \mathsf{local}(H)$, either there exists a non-local $\beta$ not located in an aborted or live transaction such that $\beta <_{\mathsf{wr}_x(H)} \alpha'$, or there is no such $\beta$ and $v = v_{\mathsf{init}}$.

We also say that a history $H$ is *consistent*, written $\mathsf{cons}(H)$, if all of its matching read requests and responses are.

We now present the definition of an opacity graph of a history with mixed transactional/non-transactional accesses.

**Definition 6.3.** The *opacity graph* of a history $H$ is a tuple $G = (N, \mathsf{vis}, \mathsf{HB}, \mathsf{WR}, \mathsf{WW}, \mathsf{RW})$, where:
- $N = \mathsf{txns}(H) \cup \mathsf{nontxn}(H)$ is the set of nodes.
- $\mathsf{vis} \subseteq N$ is a *visibility* predicate, such that it holds of all non-transactional accesses and committed transactions and does not hold of all aborted and live transactions.
- $\mathsf{HB} \in \mathcal{P}(N \times N)$ is such that

$$n \xrightarrow{\mathsf{HB}} n' \iff \exists \alpha \in n, \alpha' \in n'.\ \alpha <_{\mathsf{hb}(H)} \alpha'.$$

- $\mathsf{WR} \in \mathsf{Reg} \to \mathcal{P}(N \times N)$ specifies *read-dependency* relations on nodes: for each $x \in \mathsf{Reg}$,

$$n \xrightarrow{\mathsf{WR}_x} n' \iff n \ne n' \wedge \exists \alpha \in n, \alpha' \in n'.\ \alpha <_{\mathsf{wr}_x(H)} \alpha',$$

where the relation on actions $\mathsf{wr}_x(H)$ is defined in §3. We require that each node that is read from be visible:

$$\forall n, x.\ n \xrightarrow{\mathsf{WR}_x} \_ \implies \mathsf{vis}(n).$$

- WW ∈ Reg → $\mathcal{P}(N \times N)$ specifies *write-dependency* relations, such that for each $x \in$ Reg, $WW_x$ is an irreflexive total order on $\{n \in N \mid \text{vis}(n) \wedge (\_,\_,\text{write}(x,\_)) \in n\}$.
- RW ∈ Reg → $\mathcal{P}(N \times N)$ specifies *anti-dependency* relations, computed from WR and WW as follows:

$$n \xrightarrow{RW_x} n' \iff n \neq n' \wedge ((\exists n''. n'' \xrightarrow{WW_x} n' \wedge n'' \xrightarrow{WR_x} n) \\ \vee (\text{vis}(n') \wedge (\_,\_,\text{write}(x,\_)) \in n' \\ \wedge (\_,\_,\text{ret}(x, v_{\text{init}})) \in n)).$$

We let Graph(H) denote the set of all opacity graphs of $H$. We say that a graph $G$ is *acyclic*, written acyclic($G$), if edges from HB, WR, WW and RW do not form a cycle.

The nodes in our opacity graph include transactions and non-transactional accesses in $H$. The intention of the vis predicate is to mark those nodes that have taken effect, in particular, commit-pending transactions that should be considered committed (cf. history completions in §2.4). The other components, intuitively, constrain the order in which the nodes should go in a sequential history witnessing the strong opacity of $H$. The HB relation is the lifting of happens-before to the nodes of the graph. A read-dependency $n \xrightarrow{WR_x} n'$ specifies when the node $n'$ reads a value of $x$ written by another node $n$. A write-dependency $n \xrightarrow{WW_x} n'$ specifies when $n'$ overwrites a value of $x$ written by $n$; for the writes to take effect, both nodes should be visible. Finally, an anti-dependency $n \xrightarrow{RW_x} n'$ specifies when $n$ reads a value of $x$ overwritten by $n'$; the initial value $v_{\text{init}}$ of $x$ is considered overwritten by any write to the register.

The following lemma (proved in §B.2) shows that we can check strong opacity of a history by checking its consistency and the acyclicity of its opacity graph. Then the theorem following from it gives a criterion for the strong opacity of a TM $\mathcal{H}$.

**Lemma 6.4.** $\forall H. (\text{cons}(H) \wedge \exists G \in \text{Graph}(H). \text{acyclic}(G)) \\ \implies (\exists H' \in \mathcal{H}_{\text{atomic}}. H \sqsubseteq H').$

**Theorem 6.5.** $\mathcal{H} \sqsubseteq \mathcal{H}_{\text{atomic}}$ holds, if the following is true:

$\forall H \in \mathcal{H}. \text{cons}(H) \wedge \exists G \in \text{Graph}(H). \text{acyclic}(G).$

In comparison to the graph characterization of the usual opacity [20] for TMs without mixed transactional/non-transactional accesses, ours is more complex: the graph includes non-transactional accesses and the acyclicity check has to take into account the happens-before relation. We now show that, to prove the strong opacity of a TM using Theorem 6.5, we need to make only a minimal adjustment to a proof of its usual opacity using graph characterization. The latter characterization includes only transactions as nodes of the graph, and instead of happens-before, considers the lifting of the real-time order from §4 to transactions: for a history $H$, we let RT($H$) be the relation between transactions in $H$ such that $T <_{\text{RT}(H)} T'$ iff for some $\alpha \in T$ and $\alpha' \in T'$ we have $\alpha <_{\text{rt}(H)} \alpha'$.

In the following we abuse notation and denote by WR also the relation $\bigcup_{x \in \text{Reg}} WR_x$, and similarly for WW and RW.

**Theorem 6.6.** *Let a history* $H \in \mathcal{H}_{\mathbb{C}}|_{\text{DRF}}$ *and an opacity graph* $G = (N, \text{vis}, \text{HB}, \text{WR}, \text{WW}, \text{RW}) \in \text{Graph}(H)$ *be such that the relation* (HB ; (WR ∪ WW ∪ RW)) *is irreflexive. If G contains a cycle, then it also contains a cycle over transactions only with edges from* RT ∪ WR ∪ WW ∪ RW.

Thus, the theorem allows us to modularize the proof of the acyclicity of an opacity graph into: (i) checking the absence of "small" cycles with a single dependency edge; and (ii) checking the absence of cycles in the projection of the graph to transactions, with real-time order replacing happens-before. The latter acyclicity check is exactly the one required in the graph characterization of the usual opacity [20]. In the next section, we show how the theorem enables a simple proof of strong opacity of a realistic TM, TL2 [12].

The proof of the Theorem 6.6 is given in §B.3. Its main idea lies in the observation that any edge in WR ∪ WW ∪ RW, where one of the endpoints is a transaction and one is a non-transactional access, yields a pair of conflicting actions in $H$. Since $H$ is DRF, this means that the nodes are related by HB one way or another, and the irreflexivity of (HB ; (WR ∪ WW ∪ RW)) means that the dependency edge has to be covered by HB. Using this, we can transform any cycle in the graph into one in RT ∪ WR ∪ WW ∪ RW by replacing segments of edges involving non-transactional accesses by the real-time order.

## 7 Case Study: TL2

*The TL2 algorithm.* The metadata maintained by the TL2 software TM are summarized in Figure 7. For each register $x$, TL2 maintains its value reg[$x$], version number ver[$x$] and a write-lock lock[$x$]. New version numbers are generated with the help of a global counter clock, which transactions advance on commit. For every thread $t$, the TM maintains a flag active[$t$], which indicates that the thread $t$ is currently performing a transaction and is used to implement fences. TL2 also maintains metadata for each transaction $T$: a read-set rset[$T$] of registers $T$ has read from, a write-set wset[$T$] of registers and values $T$ intends to write to.

For brevity, we only provide pseudocode for transaction commits and fences, and describe the initialization, read, and write informally. When a transaction $T$ starts in a thread $t$, it sets the flag active[$t$] to true, and stores the value of clock into a local variable rver[$T$], which determines $T$'s *read timestamp*: TL2 allows $T$ to read registers only with versions less than or equal to rver[$T$]. The write of a value $v$ into a register $x$ simply adds the pair $(x, v)$ to the write-set wset[$T$].

Each time $T$ performs a read from a register $x$, it first checks if it has already performed a write to $x$, in which case it returns that the value for $x$ from the write-set wset[$T$]. In other cases, $T$ reads the current value reg[$x$] and checks that its version is less than or equal to rver[$T$]; if not, TL2 aborts the transaction.

Upon a commit, the current transaction $T$ executes the function txcommit in Figure 7. The commit starts by acquiring locks on each register in the write-set wset[$T$] (lines 11–18). Next, $T$ fetches-and-increments the value of clock, which it stores into wver[$T$] and uses as the version for the new values $T$ will write to registers—its *write timestamp* (line 19). Afterwards, $T$ ensures that each register $x$ in rset[$T$] has not been modified during the execution of $T$ by checking that $x$'s version ver[$x$] remains less than or equal to rver[$T$] and that $x$ is not currently locked (lines 20–26). The transaction then proceeds to write to the registers and release the locks one register at a time (lines 27–30). Finally, $T$ commits. Upon aborting or committing at lines 18, 26 or 31, $T$ executes a handler that clears the active[$t$] flag (not shown in the code).

We consider a simple implementation of transactional fences in lines 33–39 (taken from [17]). The implementation works in two steps: it first determines which transactions the fence should wait for by checking and storing their active flags, and then blocks until the threads performing those transactions clear their active flags.

***Proof overview.*** Due to space constraints, we only give an overview of the proof of the strong opacity of TL2. To generate the set $\mathcal{H}_{TL2}$ of all histories of TL2, we consider the *most general client* of TL2: a program where every thread non-deterministically chooses the commands to execute. The well-formedness conditions on fences from Definition 2.1 can be established with a simple reasoning about the fence function in lines 33–39 independently from the rest of the proof. We prove strong opacity using Theorem 6.5: for every execution of the most general client of TL2 with a DRF history $H$, we show that $H$ is consistent and build an opacity graph. To this end, we only need to define a visibility relation vis and write-dependencies WW, as the other components of the graph can be computed from these and $H$.

The consistency proof and the construction of the graph are inductive in the length of the execution of the most general client. We start with an empty trace, an empty history and an empty graph, and extend them as the executions proceeds. Whenever a non-transactional access $v$ is executed, we add a new visible node to the graph. When $v$ is a write to a register $x$, we also append it to the total order WW$_x$. Whenever a new transaction $T$ starts, we add a corresponding invisible node. When $T$ executes the txcommit function, if it reaches line 27, then we are sure $T$ is going to commit. At this point we therefore we make $T$ visible and append it to the total order WW$_x$ for each register $x \in$ wset[$T$].

We need to show that, whenever the graph is extended with new edges, it stays acyclic. To this end, we use Theorem 6.6 to reduce the acyclicity check to the one required when proving the usual opacity, i.e., checking the absence

```
1  Value clock, reg[NRegs], ver[NRegs];
2  Lock lock[NRegs];
3  Bool active[NThreads];
4  Set<Register> rset; // for each transaction
5  Map<Register, Value> wset; // for each transaction
6  Value rver; // for each transaction, initially ⊥
7  Value wver; // for each transaction, initially ⊤
8
9  function txcommit(Transaction T):
10    Set<Lock> lset := ∅;
11    foreach x in wset[T]:
12      Bool locked := lock[x].trylock();
13      if (¬locked):
14        lset.add(x);
15      else:
16        foreach y in lset[T]:
17          lock[y].unlock();
18        return aborted(T);
19    wver[T] := fetch_and_increment(clock)+1;
20    foreach x in rset[T].keys():
21      Bool locked := lock[x].test();
22      Value ts := ver[x];
23      if (locked ∨ rver[T] < ts):
24        foreach y in lset[T]:
25          lock[y].unlock();
26        return aborted(T);
27    foreach (x, v) in wset[T]:
28      reg[x] := v;
29      ver[x] := wver[T];
30      lock[x].unlock();
31    return committed(T);
32
33  function fence():
34    Bool r[NThreads]; // initially all false
35    foreach t in ThreadID:
36      r[t] := active[t];
37    foreach t in ThreadID:
38      if (r[t]):
39        while (active[t]);
40    return;
```

**Figure 7.** A fragment of the TL2 algorithm.

of cycles over transactions in RT ∪ WR ∪ WW ∪ RW (checking the absence of cycles with a single dependency is easier and we omit its description for brevity). In our proof, only graph updates of the read and commit operations of each transaction impose proof obligations.

At every step of the graph construction, we maintain an inductive invariant that helps us prove both consistency of the history and the acyclicity of RT ∪ WR ∪ WW ∪ RW. Its most important part associates a notion of time with the edges of the graph based on the read and write timestamps of transactions:

1. $\forall T, T'. T \xrightarrow{\text{RT}} T' \implies \text{rver}[T'] = \bot \vee$
   $((\text{vis}(T) \implies \text{wver}[T] \leq \text{rver}[T']) \wedge$
   $(\neg\text{vis}(T) \implies \text{rver}[T] \leq \text{rver}[T'])).$
2. $\forall T, T'. T \xrightarrow{\text{WR}} T' \implies \text{wver}[T] \leq \text{rver}[T'].$
3. $\forall T, T'. T \xrightarrow{\text{RW}} T' \implies \text{rver}[T] < \text{wver}[T'].$
4. $\forall T, T'. T \xrightarrow{\text{WW}} T' \implies \text{wver}[T] < \text{wver}[T'].$

Property 1 asserts that, whenever a transaction $T'$ occurs after a completed transaction $T$ in the real time, it either has not yet generated a read timestamp $\text{rver}[T']$, or it has and $\text{rver}[T']$ is greater or equal to $\text{wver}[T]$ (when $T$ is visible and, therefore, committed) or $\text{rver}[T]$ (otherwise). Property 2 asserts that, whenever a transaction $T'$ reads a value of a register written by a transaction $T$, the version that $T'$ assigned to the register may not be greater than the read timestamp of $T$. This is validated by the check TL2 performs when reading registers. Property 3 asserts that a transaction $T'$ overwriting the value read by a transaction $T$ has the write timestamp greater than the read timestamp of $T$. It holds because, if $T'$ commits its write after $T$ reads the previous value of the register, then $T$ generates its read timestamp before $T'$ generates its write timestamp. Property 4 follows from the mutual exclusion that TL2 ensures for committing transactions that write the same register $x$ (using $\text{lock}[x]$). Since writes in commit operations occur within a critical section, write dependencies are always consistent with the order on write timestamps.

With the help of the above invariant, we establish that for a path between any transactions $T$ and $T'$ in the graph, certain inequalities between their timestamps take place depending on visibility of the two transactions, such as the following:

$$\text{vis}(T) \wedge \text{vis}(T') \implies \text{wver}[T] < \text{wver}[T']. \quad (1)$$

Using this and other minor observations, we can demonstrate that graph updates preserve the acyclicity of RT∪WR∪WW∪RW, by showing that a cycle would imply a contradiction involving the timestamps of transactions. As an example of such reasoning, consider a transaction $T$ executing the txcommit operation. As $T$ reaches line 27, we mark it as visible and add new write dependencies in the graph. Let us assume that adding $T' \xrightarrow{\text{WW}} T$, where $T'$ is some transaction, causes a cycle over transactions. Then there must exist a path from $T$ to $T'$. Note that $\text{vis}(T)$ and $\text{vis}(T')$ both hold, since they are ordered by WW. By (1), $\text{wver}[T] < \text{wver}[T']$ holds, because there is a path from $T$ to $T'$. On the other hand, Property 4 above gives us $\text{wver}[T'] < \text{wver}[T]$, since $T' \xrightarrow{\text{WW}} T$ is in the graph. Thus, we have arrived to a contradiction.

## 8 Related and Future Work

In this paper we have concentrated on one technique for ensuring privatization safety—transactional fences. However, there have been several proposals of alternative techniques (see [21, §4.6.1] for a survey), and in the future, we plan to address these. In particular, some TMs do not require transactional fences for safe privatization [10, 13, 36, 41], even though the programmer still has to follow a certain DRF discipline. Such a discipline has been proposed by Dalessandro and Scott [8, 9], but it did not come with a formal justification, such as our proofs of the Fundamental Property and TM correctness.

Kestor et al. [24] proposed a notion of DRF for TMs that do not support safe privatization and a race-detection tool for this notion. Unlike us, they do not consider transactional fences, so that the only way to safely privatize an object is to agree on its status outside transactions (Figure 6). Our notion of DRF specializes to the one by Kestor et al. if we consider only histories without fences. We hope that, in the future, race-detection tools like the one of Kestor et al. can be adapted to detect our notion of data races.

Lesani et al. [28] proposed a transactional DRF based on TMS [14], a TM consistency criterion. However, as they acknowledge, their proposal does not support privatization.

To the best of our knowledge, a line of work by Abadi et al. was the only one that proposed disciplines for privatization with a formal justification of their safety [3, 4]. However, they did not take into account transactional fences and considered programming disciplines more restrictive than ours. Their *static separation* [4] ensures strong atomicity by not mixing transactional and non-transactional accesses to the same register. *Dynamic separation* [3] relaxes this by introducing explicit commands to privatize and publish an object. We believe such disciplines are particular ways of achieving the more general notion of data-race freedom that we adopted.

We have previously proposed a logic for reasoning about programs using RCU [17]. Since transactional fences are similar to RCU, we believe this logic can be adapted to guide programmers in inserting fences to satisfy our notion of DRF.

In this paper we assumed sequential consistency as a baseline non-transactional memory model. However, transactions are being integrated into languages, such as C++, that have weaker memory models [1]. Our definition of a data race is given in the axiomatic style used in the C++ memory model [2]. For this reason, we believe that our results can in the future be adapted to the more complex setting of C++.

Guerraoui et al. [18] considered TMs that provide strong atomicity without making any assumptions about the client program. They formalized the requirement on such TMs as parameterized opacity and proved the impossibility of achieving it on many memory models without instrumenting non-transactional accesses. This result justifies our decision to provide strong atomicity only to DRF programs.

*Acknowledgments.* We thank Michael Spear and Tingzhe Zhou for discussions about privatization in TM.

## A Programming Language
### A.1 Formal Definition of Traces

To formalize restrictions on accesses to variables by primitive commands, we partition the set PComm into $m$ classes: $\text{PComm} = \biguplus_{t=1}^{m} \text{LPcomm}_t$. The intention is that commands from $\text{LPcomm}_t$ can access only the local variables of thread $t$ ($\text{LVar}_t$). To ensure that in our programming language a thread $t$ does not access local variables of other threads, we require that the thread cannot mention such variables in the conditions of if and while commands and can only use primitive commands from $\text{LPcomm}_t$.

**Definition A.1.** A *trace* $\tau$ is a finite sequence of actions satisfying the following well-formedness conditions:

1. every action in $\tau$ has a unique identifier: if $\tau = \tau_1\,(a_1,\_,\_)\,\tau_2\,(a_2,\_,\_)\,\tau_3$ then $a_1 \neq a_2$;
2. commands in actions executed by a thread $t$ do not access local variables of other threads $t' \neq t$: if $\tau = \_\,(\_,t,c)\,\_$ then $c \in \text{LPcomm}_t$;
3. every write operation writes a unique value: if $\tau = \_\,(\_,\_,\text{write}(\_,v))\,\_\,(\_,\_,\text{write}(\_,v'))\,\_$ then $v \neq v'$;
4. for every thread $t$, the projection $\tau|_t$ of $\tau$ onto the actions by $t$ cannot contain a request action immediately followed by a primitive action: if $\tau|_t = \_\alpha_1\alpha_2\_$ and $\alpha_1$ is a request then $\alpha_2$ is a response;
5. request and response actions are properly matched: for every thread $t$, $\text{history}(\tau)|_t$ consists of alternating request and corresponding response actions, starting from a request action;
6. actions denoting the beginning and end of transactions are properly matched: for every thread $t$, in the projection of $\tau|_t$ to txbegin, committed and aborted actions, txbegin alternates with committed or aborted, starting from txbegin;
7. non-transactional accesses execute atomically: if $\tau = \tau_1\,\alpha\,\tau_2$, where $\alpha$ is a read or a write request action by thread $t$, and all the transactions of $t$ in $\tau_1$ completed, then $\tau_2$ begins with a response to $\alpha$.
8. non-transactional accesses never abort: if $\tau = \_\,\alpha_1\,\alpha_2\,\tau_2$, where $\alpha_1$ is a non-transactional request action then $\alpha_2$ is not an aborted action;
9. fence actions may not occur inside transactions; if $\tau = \tau_1\,(t,\text{fbegin})\,\_$ the all the transactions $t$ in $\tau_1$ completed; and
10. fence blocks until all active transactions complete: if $\tau = \tau_1\,(\_,t,\text{txbegin})\,\tau_2\,(\_,t',\text{fbegin})\,\tau_3\,(\_,t',\text{fend})\,\tau_4$ then either $\tau_2$ or $\tau_3$ contains an action of the form $(\_,t,\text{committed})$ or $(\_,t,\text{aborted})$.

### A.2 Formal Definition of the Semantics of the Programming Language

This section formally defines the set $[\![P]\!](s)$. It is computed in two stages. First, we compute a set $A(P)$ of traces that resolves all issues regarding sequential control flow and interleaving. Intuitively, if one thinks of each thread $C_t$ in $P$ as a control-flow graph, then $A(P)$ contains all possible interleavings of paths in the graphs of $C_t$, $t \in \text{ThreadID}$ starting from their initial nodes. The set $A(P)$ is a superset of all the traces that can actually be executed: e.g., if a thread executes the command "$x := 1;\, \text{if}\,(x = 1)\, y := 1\, \text{else}\, y := 2$" where $x, y$ are local variables, then $A(P)$ will contain a trace where $y := 2$ is executed instead of $y := 1$. To filter out such nonsensical traces, we *evaluate* every trace to determine whether it is **valid**, i.e., whether its control flow is consistent with the effect of its actions on program variables. This is formalized by a function $\text{eval} : \text{State} \times \text{Trace} \to \mathcal{P}(\text{State}) \cup \{\frac{\iota}{\iota}\}$ that, given an initial state and a trace, produces the set of states resulting from executing the actions in the trace, an empty set if the trace is invalid, or a special state $\frac{\iota}{\iota}$ if the trace contains a fault action. Thus, $[\![P]\!](s) = \{\tau \in A(P) \mid \text{eval}(s, \tau) \neq \emptyset\}$.

When defining the semantics, we encode the evaluation of conditions in if and while statements with assume commands. More specifically, we expect that the sets $\text{LPcomm}_t$ contain special primitive commands $\text{assume}(b)$, where $b$ is a Boolean expression over local variables of thread $t$, defining the condition. We state their semantics formally below; informally, $\text{assume}(b)$ does nothing if $b$ holds in the current program state, and stops the computation otherwise. Thus, it allows the computation to proceed only if $b$ holds. The assume commands are only used in defining the semantics of the programming language; hence, we forbid threads from using them directly.

**The trace set** $A(P)$**.** The function $A'(\cdot)$ in Figure 8 maps commands and programs to sequences of actions they may produce. Technically, $A'(\cdot)$ might contain sequences that are not traces, e.g., because they do not have unique identifiers or continue beyond a fault command. This is resolved by intersecting the set $A'(P)$ with the set of all traces to define $A(P)$. $A'(C)t$ gives the set of action sequences produced by a command $C$ when it is executed by thread $t$. To define $A'(P)$, we first compute the set of all the interleavings of action sequences produced by the threads constituting $P$. Formally, $\tau \in \text{interleave}(\tau_1, \ldots, \tau_m)$ if and only if every action in $\tau$ is performed by some thread $t \in \{1, \ldots, m\}$, and $\tau|_t = \tau_t$ for every thread $t \in \{1, \ldots, m\}$. We then let $A'(P)$ be the set of all prefixes of the resulting sequences which respect Definition A.1, as denoted by the prefix operator. We take prefix closure here (while respecting the atomicity of non transactional access) to account for incomplete program computations as well as those in which the scheduler preempts a thread forever.

$A'(c)t$ returns a singleton set with the action corresponding to the primitive command $c$ (primitive commands execute atomically). $A'(C_1; C_2)t$ concatenates all possible action sequences corresponding to $C_1$ with those corresponding to $C_2$. The set of action sequences of a conditional considers

$$
\begin{aligned}
A'(c)t &= \{(\_, t, c)\} \\
A'(C_1; C_2)t &= \{\tau_1 \tau_2 \mid \tau_1 \in A'(C_1)t \land \tau_2 \in A'(C_2)t\} \\
A'(\text{if } (b) \text{ then } C_1 \text{ else } C_2)t &= \{(\_, t, \text{assume}(b)) \, \tau_1 \mid \tau_1 \in A'(C_1)t\} \cup \{(\_, t, \text{assume}(\neg b)) \, \tau_2 \mid \tau_2 \in A'(C_2)t\} \\
A'(\text{while } (b) \text{ do } C)t &= \{\tau_1 \tau_2 \ldots \tau_{2n} (\_, t, \text{assume}(\neg b)) \mid n \in \mathbb{N} \land \forall j \in \{1, \ldots, n\}. \, \tau_{2j-1} = (\_, t, \text{assume}(b)) \\
&\qquad \land \tau_{2j} \in A'(C)t\} \cup \{(\_, t, \text{assume}(\neg b))\} \\
A'(l := x.\text{read}())t &= \{(\_, t, \text{read}(x)) \, (\_, t, \text{ret}(v)) \, (\_, t, l := v) \mid v \in \mathbb{Z}\} \cup \{(\_, t, \text{read}(x)) \, (\_, t, \text{aborted})\} \\
A'(x.\text{write}(e))t &= \{(\_, t, \text{assume}(e = v)) \, (\_, t, \text{write}(x, v)) \, (\_, t, \text{ret}(\bot))) \mid v \in \mathbb{Z}\} \\
&\quad \cup \{(\_, t, \text{assume}(e = v)) \, (\_, t, \text{write}(x, v)) \, (\_, t, \text{aborted}) \mid v \in \mathbb{Z}\} \\
A'(\text{fence})t &= \{(\_, t, \text{fbegin}) \, (\_, t, \text{fend})\} \\
A'(x := \text{atomic } \{C\})t &= \{(\_, t, \text{txbegin}) \, (\_, t, \text{aborted}) \, (\_, t, x := \text{aborted})\} \\
&\quad \cup \{(\_, t, \text{txbegin}) \, (\_, t, \text{ok}) \, \tau \, (\_, t, \text{aborted}) \, (\_, t, x := \text{aborted}) \mid \\
&\qquad \tau \, (\_, t, \text{aborted}) \, \tau' \in A'(C)t \land (\_, t, \text{aborted}) \notin \tau\} \\
&\quad \cup \{(\_, t, \text{txbegin}) \, (\_, t, \text{ok}) \, \tau \, (\_, t, \text{txcommit}) \, (\_, t, r) \, (\_, t, x := r) \mid \\
&\qquad \tau \in A'(C)t \land (\_, t, \text{aborted}) \notin \tau \land (r = \text{committed} \lor r = \text{aborted})\} \\
A'(C_1 \parallel \ldots \parallel C_m) &= \text{prefix}(\bigcup \{\text{interleave}(\tau_1, \ldots, \tau_m) \mid \forall t. \, 1 \leq t \leq m \implies \tau_t \in A'(C_t)t\}) \\
A(P) &= A'(P) \cap \text{Trace}
\end{aligned}
$$

**Figure 8.** The definition of $A(P)$.

cases where either branch is taken. We record the decision using an assume action; at the evaluation stage, this allows us to ensure that this decision is consistent with the program state. The set of action sequences for a loop is defined by considering all possible unfoldings of the loop body. Again, we record branching decisions using assume actions.

The set of action sequences of read and write accesses includes both sequences where the access executes successfully and where the current transaction is aborted. The former set is constructed by nondeterministically choosing an integers $v$ to describe the the return and parameter for the read and write accesses, respectively. To ensure that $e$ indeed evaluates to $v$, in the case of a write, Note that some of the choices here might not be feasible: the chosen $v$ might not be the value of the parameter expression $e$ when the method is invoked. Such infeasible choices are filtered out at the following stages of the semantics definition: the former in the definition of $[\![P]\!](s)$ by the use of evaluation and the semantics of assume, and the latter in the definition of $[\![P, \mathcal{H}]\!](s)$ by selecting the sequences from $[\![P]\!](s)$ that interact with the transactional memory correctly. The set of action sequences of a fence command is comprised of all traces comprised of a fbegin action followed by a fend action, indicating that once a fence command is invoked, the thread gets blocked until it ends. The set of action sequences of $x := \text{atomic } \{C\}$ contains those in which $C$ is aborted in the middle of its execution (at an object operation or right after it begins) and those in which $C$ executes until completion and then the transaction commits or aborts.

**Semantics of primitive commands.** To define evaluation, we assume a semantics of every command $c \in \text{PComm}$, given by a function $[\![c]\!]$ that defines how the program state is transformed by executing $c$. As we noted before, different classes of primitive commands are supposed to access only certain subsets of variables. To ensure that this is indeed the case, we define $[\![c]\!]$ as a function of only those variables that $c$ is allowed to access. Namely, the semantics of $c \in \text{LPcomm}_t$ is given by

$$[\![c]\!] : (\text{LVar}_t \to \mathbb{Z}) \to \mathcal{P}(\text{LVar}_t \to \mathbb{Z}).$$

Note that we allow $c$ to be non-deterministic.

For a valuation $q$ of variables that $c$ is allowed to access, $[\![c]\!](q)$ yields the set of their valuations that can be obtained by executing $c$ from a state with variable values $q$. For example, an assignment command $l := l'$ has the following semantics:

$$[\![l := l']\!](q) = \{q[l \mapsto q(l')]\}.$$

We define the semantics of assume commands following the informal explanation given at the beginning of this section: for example,

$$[\![\text{assume}(l = v)]\!](q) = \begin{cases} \{q\}, & \text{if } q(l) = v; \\ \emptyset, & \text{otherwise.} \end{cases} \quad (2)$$

Thus, when the condition in assume does not hold of $q$, the command stops the computation by not producing any output.

We lift functions $[\![c]\!]$ to full states by keeping the variables that $c$ is not allowed to access unmodified and producing ↯

if $c$ faults. For example, if $c \in \text{LPcomm}_t$, then

$$[\![c]\!](s) = \{s|_{\text{LVar}\setminus\text{LVar}_t} \uplus q \mid q \in [\![c]\!](s|_{\text{LVar}_t})\},$$

where $s|_V$ is the restriction of $s$ to variables in $V$. (For simplicity, we assume commands to not fault.)

**Trace evaluation.** Using the semantics of primitive commands, we first define the evaluation of a single action on a given state:

$$\text{eval} : \text{State} \times \text{Action} \to \mathcal{P}(\text{State})$$
$$\text{eval}(s, (\_, t, c)) = [\![c]\!](s);$$
$$\text{eval}(s, \psi) = \{s\}.$$

Note that this does not change the state $s$ as a result of TM interface or fence actions, since their return values are assigned to local variables by separate actions introduced when generating $A(P)$. We then lift eval to traces as follows:

$$\text{eval} : \text{State} \times \text{Trace} \to \mathcal{P}(\text{State})$$
$$\text{eval}(s, \tau) = \begin{cases} \emptyset, & \text{if } \tau = \tau'\varphi \land \text{eval}(s, \tau') = \emptyset; \\ \text{evalna}(s, \tau|_{\neg\text{abortact}}), & \text{otherwise,} \end{cases}$$

where $\tau|_{\neg\text{abortact}}$ denotes the trace obtained from $\tau$ by removing all actions inside aborted transactions, and

$$\text{evalna}(s, \tau) =$$
$$\begin{cases} \{s\}, & \text{if } \tau = \varepsilon; \\ \{s'' \in \text{eval}(s', \varphi) \mid s' \in \text{evalna}(s, \tau')\}, & \text{if } \tau = \tau'\varphi. \end{cases}$$

The set of states resulting from evaluating trace $\tau$ from state $s$ is effectively computed by the helper function $\text{evalna}(s, \tau)$, which ignores actions inside aborted transactions to model local variable roll-back. However, ignoring the contents of aborted transactions completely poses a risk that we might consider traces including sequences of actions inside aborted transactions that yield an empty set of states. To mitigate this, $\text{eval}(s, \tau)$ recursively evaluates every prefix of $\tau$, thus ensuring that sequences of actions inside aborted transaction are valid.

Recall that we define $[\![P]\!](s)$ as the set of those traces from $A(P)$ that can be evaluated from $s$ without getting stuck, as formalized by eval. Note that this definition enables the semantics of assume defined by (2) to filter out traces that make branching decisions inconsistent with the program state. For example, consider again the program "$l := 1$; if $(l = 1)$ $l' := 1$ else $l' := 2$". The set $A(P)$ includes traces where both branches are explored. However, due to the semantics of the assume actions added to the traces according to Figure 8, only the trace executing $l' := 1$ will result in a nonempty set of final states after the evaluation and, therefore, only this trace will be included into $[\![P]\!](s)$.

## B  Proofs
### B.1  Proof of Lemma 5.4
**Lemma 5.4.** *If $\mathcal{H}$ is a TM such that $\mathcal{H}|_{\text{DRF}} \sqsubseteq \mathcal{H}_{\text{atomic}}$, then:*

1. $\forall P, s. \text{DRF}(P, s, \mathcal{H}) \implies [\![P]\!](\mathcal{H}, s) \leq [\![P]\!](\mathcal{H}_{\text{atomic}}, s)$.
2. $\forall P, s. \text{DRF}(P, s, \mathcal{H}_{\text{atomic}}) \implies \text{DRF}(P, s, \mathcal{H})$.

The key step in the proof of Lemma 5.4(1) is the next lemma. It shows that a trace $\tau_H$ with a history $H$ can be transformed into an equivalent trace $\tau_S$ with a history $S$ that is in the opacity relation with $H$.

**Lemma B.1** (Rearrangement).

$$\forall H, S \in \text{History}. H \sqsubseteq S \implies (\forall \tau_H. \text{history}(\tau_H) = H \implies$$
$$\exists \tau_S. \text{history}(\tau_S) = S \land \tau_H \sim \tau_S).$$

The proof follows the proof of the corresponding lemma in [7] and is omitted.

We also rely on the following proposition that allows us to conclude that the trace $\tau_S$ resulting from the rearrangement in Lemma B.1 can be produced by a program $P$ if so can the original trace $\tau_H$.

**Proposition B.2.** *If $\tau_H \in [\![P]\!](s)$ and $\tau_H \sim \tau_S$, then $\tau_S \in [\![P]\!](s)$.*

This proposition follows immediately from Definition 5.1 and the definition of $A(P)$ in Figure 8.

Lemma B.1 and Proposition B.2 together yield the following corollary, from which Lemma 5.4(1) follows.

**Corollary B.3.** *If $\mathcal{H}$ is a TM such that $\mathcal{H}|_{\text{DRF}} \sqsubseteq \mathcal{H}_{\text{atomic}}$, then:*

$$\forall P, s, \tau_H. \tau_H \in [\![P]\!](\mathcal{H}, s) \land \text{DRF}(\tau_H) \implies$$
$$\exists \tau_S. \tau_S \in [\![P]\!](\mathcal{H}_{\text{atomic}}, s) \land$$
$$\text{history}(\tau_H) \sqsubseteq \text{history}(\tau_S) \land \tau_H \sim \tau_S.$$

The following proposition states the prefix-closure property of the programming language semantics.

**Proposition B.4.** *For every program $P$, TM $\mathcal{H}$ and state $s$, the set $[\![P]\!](\mathcal{H}, s)$ is prefix-closed.*

*Proof of Lemma 5.4(2).* Let us consider a TM $\mathcal{H}$ such that $\mathcal{H}|_{\text{DRF}} \sqsubseteq \mathcal{H}_{\text{atomic}}$, any client program $P$ and an initial state $s$. We prove the theorem by contrapositive, i.e., by demonstrating:

$$\neg\text{DRF}(P, s, \mathcal{H}) \implies \neg\text{DRF}(P, s, \mathcal{H}_{\text{atomic}})$$

Let us assume that $\text{DRF}(P, s, \mathcal{H})$ does not hold. By Definition 3.3, there exists $\hat{\tau} \in [\![P]\!](\mathcal{H}, s)$ such that its history $\hat{H} = \text{history}(\hat{\tau})$ is racy, i.e., $\text{DRF}(\hat{H})$ does not hold. More specifically, $\hat{\tau}$ contains a data race $(\alpha', \alpha)$. Note that there might be multiple races in $\hat{\tau}$, and we choose such race $(\alpha', \alpha)$ that the later of the two actions, $\alpha$, is the earliest in the execution order of $\hat{H}$. We split the further proof depending on whether $\alpha$ is transactional or non-transactional.

Case 1: $\alpha$ is non-transactional. Let $\hat{\tau}$ take form of $\hat{\tau} = \tau\alpha\beta\_$, where $\beta$ is a matching response action for $\alpha$ and $\tau$ is a prefix of $\hat{\tau}$ containing $\alpha'$. It is easy to see that our choice of the data race is such that $\tau\alpha\beta$ is racy and $\tau$ is data-race free. Additionally, as stated in Proposition B.4, $[\![P]\!](\mathcal{H}, s)$ is prefix-closed. Therefore, both $\tau\alpha\beta \in [\![P]\!](\mathcal{H}, s)$ and $\tau \in [\![P]\!](\mathcal{H}, s)$ hold. By Corollary B.3, there exists a trace $\tau' \in [\![P]\!](\mathcal{H}_{\text{atomic}}, s)$ such that its history $S = \text{history}(\tau') \in \mathcal{H}_{\text{atomic}}$ is in the strong opacity relation with $H = \text{history}(\tau)$ ($H \sqsubseteq S$) and $\tau \sim \tau'$.

Let us consider a trace $\tau'' = \tau'\alpha\beta'$, which extends $\tau'$ with the request action $\alpha$ and a matching response $\beta'$ in such a way that its history $S'' = S\alpha\beta' \in \mathcal{H}_{\text{atomic}}$. Such response $\beta'$ and a trace $\tau''$ exist, since $S\alpha$ is a non-interleaved history, and it is easy to see that it can always be extended with a response to $\alpha$ (not necessarily returning the same result as $\beta$).

Since $\tau \sim \tau'$, we have $\tau\alpha\beta \sim \tau'\alpha\beta$ according to Definition 5.1. Therefore, by Proposition B.2, $\tau'\alpha\beta \in [\![P]\!](s)$. Also, as evident from the definition of $A(P)$ in Figure 8, $A(P)$ is closed under replacing a return value of a trailing read-response action, so $\tau'' \in [\![P]\!](s)$ holds. Knowing that $S'' \in \mathcal{H}_{\text{atomic}}$, we conclude that $\tau'' \in [\![P]\!](\mathcal{H}_{\text{atomic}}, s)$.

We argue that $\text{hb}(H\alpha\beta) = \text{hb}(S'')$. First, since $\tau\alpha\beta \sim \tau'\alpha\beta$ and $\tau''$ differs from $\tau'\alpha\beta$ only in the return value of $\beta'$, $\text{cl}(H\alpha\beta) = \text{cl}(S'')$ and $\text{po}(H\alpha) = \text{po}(S'')$ hold. Since $\alpha$ is neither a fence action nor a beginning or an end of a transaction, $\text{af}(H\alpha\beta) = \text{af}(S'')$ and $\text{bf}(H\alpha\beta) = \text{bf}(S'')$ hold too. Finally, since $\alpha$ is non-transactional, it does not contribute to $(\text{xpo}(S'') \, ; \, \text{txwr}(S''))$. Overall, $\text{hb}(H\alpha\beta) = \text{hb}(S'')$ holds.

Knowing that $\text{hb}(\text{history}(\tau\alpha\beta)) = \text{hb}(H\alpha\beta) = \text{hb}(S'')$ and that the conflict $(\alpha', \alpha)$ is unordered in $\text{hb}(\text{history}(\tau\alpha\beta))$, we conclude that $(\alpha', \alpha)$ is a data race in $\tau''$.

CASE 2: $\alpha$ IS TRANSACTIONAL. Let $\hat{\tau}$ take form of $\hat{\tau} = \tau\alpha\_$, where the prefix $\tau$ contains $\alpha'$. It is easy to see that our choice of the data race is such that $\tau\alpha$ is racy and $\tau$ is data-race free. Additionally, as stated in Proposition B.4, $[\![P]\!](\mathcal{H}, s)$ is prefix-closed. Therefore, both $\tau\alpha \in [\![P]\!](\mathcal{H}, s)$ and $\tau \in [\![P]\!](\mathcal{H}, s)$ hold. By Corollary B.3, there exists a trace $\tau' \in [\![P]\!](\mathcal{H}_{\text{atomic}}, s)$ such that its history $S = \text{history}(\tau') \in \mathcal{H}_{\text{atomic}}$ is in the strong opacity relation with $H = \text{history}(\tau)$ ($H \sqsubseteq S$) and $\tau \sim \tau'$.

Let us consider a trace $\tau''$ obtained from $\tau'$ by inserting $\alpha$ after the last action by the same thread. It is easy to see that $\tau''$ is well-formed and that $S' = \text{history}(\tau'')$ is non-interleaved, meaning that $S' \in \mathcal{H}_{\text{atomic}}$ holds.

Since $\tau \sim \tau'$, the following holds:

$$(\forall t \in \text{ThreadID}.\ \tau|_t = \tau'|_t) \land (\tau|_{\text{nontx}} = \tau'|_{\text{nontx}}).$$

Note that the following holds, because $\alpha$ is transactional:

$$\tau''|_{\text{nontx}} = \tau'|_{\text{nontx}} = \tau|_{\text{nontx}} = (\tau\alpha)|_{\text{nontx}}.$$

Let $t'$ be the thread of $\alpha$. For all $t \in \text{ThreadID} \setminus \{t'\}$, the following holds, because $\alpha$ is by a different thread:

$$(\tau\alpha)|_t = \tau|_t = \tau'|_t = \tau''|_t,$$

and $(\tau\alpha)|_{t'} = \tau''|_{t'}$ by construction. Overall, we have:

$$(\forall t \in \text{ThreadID}.\ (\tau\alpha)|_t = \tau''|_t) \land ((\tau\alpha)|_{\text{nontx}} = \tau''|_{\text{nontx}}),$$

so $\tau\alpha \sim \tau''$. Therefore, by Proposition B.2, $\tau'' \in [\![P]\!](s)$. Knowing that $S' \in \mathcal{H}_{\text{atomic}}$, we conclude that $\tau'' \in [\![P]\!](\mathcal{H}_{\text{atomic}}, s)$.

We argue that $\text{hb}(H\alpha) = \text{hb}(S')$. Similarly to Case 1, we observe that po, cl, af and bf components of $\text{hb}(H\alpha)$ and $\text{hb}(S')$ are equal. Let us consider the edges from $(\text{xpo}(H\alpha) \, ; \, \text{txwr}(H\alpha))$ that may appear in $\text{hb}(H\alpha) \setminus \text{hb}(H)$:

- if $\alpha$ is a write request, then $\{(\beta, \beta') \mid \beta <_{\text{xpo}(H\alpha)} \alpha <_{\text{txwr}(H\alpha)} \beta'\}$;
- if $\alpha$ is a read request, then $\{(\beta, \alpha) \mid \exists \beta'.\ \beta <_{\text{xpo}(H\alpha)} \beta' <_{\text{txwr}(H\alpha)} \alpha\}$.

Note that $\text{txwr}(H\alpha) = \text{txwr}(S')$, because $H\alpha$ and $S'$ are histories with unique writes, so read-dependencies are uniquely determined. Also, $\text{xpo}(H\alpha) = \text{xpo}(S')$, because $\alpha$ is executed within the same transaction in the two histories. Therefore, the aforementioned edges are present in $\text{hb}(H\alpha)$ if and only if they are present in $\text{hb}(S')$. Overall, $\text{hb}(H\alpha) = \text{hb}(S')$ holds.

Knowing that $\text{hb}(\text{history}(\tau\alpha)) = \text{hb}(S')$ and that the conflict $(\alpha', \alpha)$ is unordered in $\text{hb}(\text{history}(\tau\alpha))$, we conclude that $(\alpha', \alpha)$ is a data race in $\tau'\alpha$. □

### B.2 Proof of Lemma 6.4

**Lemma 6.4.** $\forall H_1.\ \text{cons}(H_1) \land (\exists G \in \text{Graph}(H_1).\ \text{acyclic}(G))$
$$\implies (\exists H_2 \in \mathcal{H}_{\text{atomic}}.\ H_1 \sqsubseteq H_2).$$

To prove Lemma 6.4, we introduce a notion of *fenced* opacity graphs, which extend opacity graphs with fence actions. We prove that fenced opacity graphs are acyclic when corresponding opacity graphs are.

Let $\text{fact}(H)$ denote the set of all fence actions of a history $H$, or formally:

$$\text{fact}(H) \triangleq \{(a, t, \text{fbegin}) \mid (a, t, \text{fbegin}) \in \text{act}(H)\}$$
$$\cup \{(a, t, \text{fbegin}) \mid (a, t, \text{fend}) \in \text{act}(H)\}.$$

**Definition B.5.** Given a history $H$ and its opacity graph $G = (N, \text{vis}, \text{HB}, \text{WR}, \text{WW}, \text{RW}) \in \text{Graph}(H)$, we define a matching fenced graph $\overline{G}$ as a tuple $(\overline{N}, \text{vis}, \overline{\text{HB}}, \text{WR}, \text{WW}, \text{RW})$, where:

- $\overline{N} = N \cup \text{fact}(H)$ extends the set of nodes of $G$ with nodes denoting fence actions of the history $H$;
- $\overline{\text{HB}} \in \mathcal{P}(\overline{N} \times \overline{N})$ is such that

$$n \xrightarrow{\overline{\text{HB}}} n' \iff \exists \alpha, \alpha'.\ (\alpha \in n \lor \alpha = n \in \text{fact}(H))$$
$$\land\ (\alpha' \in n' \lor \alpha' = n' \in \text{fact}(H)) \land \alpha <_{\text{hb}(H)} \alpha'.$$

That is, a fenced graph $\overline{G}$ corresponding to $G \in \text{Graph}(H)$ simply extends $G$ with fence actions and happens-before edges.

**Proposition B.6.** $\forall H, G.\ G \in \text{Graph}(H) \land \text{acyclic}(G)$
$$\implies \text{acyclic}(\overline{G}).$$

*Proof.* Let $G = (N, \text{vis}, \text{HB}, \text{WR}, \text{WW}, \text{RW})$ and $\overline{G} = (\overline{N}, \text{vis}, \overline{\text{HB}}, \text{WR}, \text{WW}, \text{RW})$. It is easy to see that:

$$\overline{\text{HB}} = \text{HB} \uplus \{(\alpha, \alpha') \mid \alpha <_{\text{hb}(H)} \alpha' \wedge$$
$$(\alpha \in \text{fact}(H) \vee \alpha' \in \text{fact}(H))\}$$

That is, all edges between non-fence nodes implied by transitivity via fence nodes in $\overline{G}$ are already present in $G$. We know that $G$ is acyclic. Therefore, if $\overline{G}$ contains a cycle, it necessarily involves fence actions.

Let us first consider a cycle over nodes of $\overline{G}$ consisting entirely of fence action nodes. The cycle takes the following form then:

$$\phi_1 \xrightarrow{\text{HB}} \phi_2 \xrightarrow{\text{HB}} \ldots \xrightarrow{\text{HB}} \phi_k \xrightarrow{\text{HB}} \phi_1,$$

where $k$ is the length of the cycle and $\phi_i$ is a fence action for each $i$ such that $1 \le i \le k$. For each consecutive $\phi_i$ and $\phi_{i+1}$ in the cycle, we assume that there is no non-fence action $n$ such that $\phi_i \xrightarrow{\text{HB}} n \xrightarrow{\text{HB}} \phi_{i+1}$ (we consider cycles involving non-fence actions separately). By Definition 6.3, the following must be the case:

$$\phi_1 <_{\text{hb}(H)} \phi_2 <_{\text{hb}(H)} \ldots <_{\text{hb}(H)} \phi_k <_{\text{hb}(H)} \phi_1,$$

It is easy to see that such cycle is not possible: without intermediate non-fence actions, fence actions can only be related by $\text{po}(H)$ in $\text{hb}(H)$, and $\text{po}(H)$ is not cyclic.

We now consider a cycle in $\overline{G}$ involving at least one node that does not correspond to a fence action. We consider each sequence of fence actions in the cycle surrounded by non-fence actions:

$$n \xrightarrow{\text{HB}} \phi_1 \xrightarrow{\text{HB}} \phi_2 \xrightarrow{\text{HB}} \ldots \xrightarrow{\text{HB}} \phi_k \xrightarrow{\text{HB}} n'$$

where $n$ and $n'$ denote non-fence actions (or, possibly, the same non-fence action), and for each $i$ ($1 \le i \le k$), $\phi_i$ is a fence action. Since HB is transitive, $n \xrightarrow{\text{HB}} n'$ is present in the graph $G$. By replacing each segment of fence actions in the cycle with HB-edges between non-fence actions, we transform the cycle in $\overline{G}$ into a cycle in $G$. By the premise of the proposition, $\text{acyclic}(G)$ holds. Thus, we arrive to a contradiction, meaning that $\text{acyclic}(\overline{G})$ holds. □

Before proving Lemma 6.4, let us restate the definition of $\mathcal{H}_{\text{atomic}}$ from §2.4.

**Definition B.7.** In a non-interleaved history $H$, a read request $(\_,\_,\text{read}(x))$ and its matching response $(\_,\_,\text{ret}(v))$ are said to be *legal*, if $v$ is the value returned by the last preceding $\text{write}(x, v)$ action that is not located in an aborted or live transaction different from the one of the read; if there is no such write, then $v = v_{\text{init}}$.

Thus, $\mathcal{H}_{\text{atomic}}$ can be defined as the set of all non-interleaved histories $H$ that have a completion $H^c$, in which every pair of a matching read request and response is legal.

*Proof of Lemma 6.4.* In the proof, we only consider finite histories of TM. Let $H_1$ be an consistent TM history. Assume that there exists an acyclic graph $G \in \text{Graph}(H_1)$. We consider its matching fenced graph $\overline{G} = (N, \text{vis}, \text{HB}, \text{WR}, \text{WW}, \text{RW})$, which is acyclic by Proposition B.6.

Let $k$ be the number of nodes in $N$ and a sequence $S = n_1 n_2 \ldots n_k$ be the result of a topological sort of graph $\overline{G}$. Let $\prec_S$ be a relation on nodes of $S$ such that $n \prec_S n'$ if $n$ occurs earlier than $n'$ in $S$.

We define a relation $\text{lin} \subseteq \text{act}(H_1) \times \text{act}(H_1)$ on actions of history $H_1$ so that for every actions $\alpha <_{\text{lin}} \alpha'$ if for $n, n' \in S$ such that $\alpha \in n$ and $\alpha' \in n'$, either $n = n' \wedge \alpha <_{\text{po}(H_1)} \alpha'$ or $n \ne n' \wedge n \prec_S n'$ holds. It is easy to see that lin is a linear order on $\text{act}(H_1)$.

Let $H_2$ be a sequence obtained by putting the actions of the history $H_2$ in the order lin. Note that $\text{hb}(H) \subseteq \text{lin}$. Therefore, by Definition 4.1, $H_1 \sqsubseteq H_2$ holds. It is easy to see that $H_2$ is non-interleaved. To conclude that $H_2 \in \mathcal{H}_{\text{atomic}}$, it remains to show that it has a non-interleaved completion, in which every matching read request and response are legal.

Let $H_2^c$ be a non-interleaved completion of $H_2$ obtained by committing each transaction from vis and aborting all other commit-pending transaction. We argue that $\text{cons}(H_2^c)$ holds. It is easy to see that local read actions are consistent in $H_2^c$. Consider any non-local read $(\alpha, \alpha')$. Since $\text{cons}(H_1)$ holds, there are two possibilities:

- There exists a non-local $\beta$ such that $\beta <_{\text{wr}} \alpha'$ and $\beta$ is not located in an aborted or live transaction in $H_1$. Let $n \in N$ be the node of $\hat{G}$ containing $\beta$. By Definition 6.3, $n \in \text{vis}$ holds. Therefore, $n$ is not an aborted or live transaction in $H_2^c$ either, meaning that $(\alpha, \alpha')$ is consistent in $H_2^c$.
- There is no such write $\beta$, and $\alpha'$ returns $v_{\text{init}}$. It is easy to see that $(\alpha, \alpha')$ is consistent in $H_2^c$.

We consider every read request $\alpha = (\_,\_,\text{read}(x))$ and its matching response $\alpha' = (\_,\_,\text{ret}(v))$ in $H_2^c$. When $(\alpha, \alpha') \in \text{local}(H_2^c)$, the consistency of $H_2^c$ immediately implies that $(\alpha, \alpha')$ is legal.

Let us assume that $(\alpha, \alpha') \notin \text{local}(H_2^c)$, and let $n$ be its node in the graph $\hat{G}$. By Definition 6.2, there are two possibilities:

- there exists a non-local $\beta_1$ not located in an aborted or live transaction and such that $\beta_1 <_{\text{wr}_x(H_2^c)} \alpha'$; or
- $v = v_{\text{init}}$, otherwise.

Let us assume that there is no $\beta_1$ satisfying the above and that $v = v_{\text{init}}$. Consider any action $\beta_2 = (\_,\_,\text{write}(x,\_))$ in $H_2^c$ that is not located in a live or aborted transaction, and let $n_2$ denote its node in $\hat{G}$. Knowing that $n_2$ is not an aborted transaction in $H_2^c$, we conclude that $n \in \text{vis}$ holds. By Definition 6.3, $n \xrightarrow{\text{RW}_x} n_2$ holds. Since lin is a topological sort of the order including $\text{RW}_x$, all actions of the node $n_2$ occur after all actions of the node $n$, meaning that $\beta_2$ does not precede $\alpha$ in $H_2^c$. Therefore, It is easy to see that $(\alpha, \alpha')$ is legal.

We now consider the case when in $H_2^c$ there exists a non-local $\beta_1$ not located in an aborted or live transaction and such that $\beta_1 <_{\text{wr}_x(H_2^c)} \alpha'$. It is easy to see that in order to conclude that $(\alpha, \alpha')$ is legal, we only need to prove that $\beta_1$ is the most recent write to $x$ preceding $\alpha$. Let $n$ and $n_1$ be the nodes of $\alpha$ and $\beta_1$ accordingly in the graph $\overline{G}$. Firstly, we observe that $\beta_1 <_{\text{wr}_x(H_1)} \alpha$ holds and, hence, so does $n_1 \xrightarrow{\text{WR}} n$. Since lin is a topological sort of the order including WR, actions of $n_1$ precede actions of $n$ in $H_2^c$. Therefore, $\beta_1$ precedes $\alpha$ in $H_2^c$. Let us consider any other write $\beta_2 = (\_, \_, \text{write}(x, \_))$ that occurs after $\beta_1$ in $H_2^c$ and that is not located in a live or aborted transaction. Let $n_2$ be its node in the graph $\overline{G}$. Knowing that $n_2$ is not an aborted transaction in $H_2^c$, we conclude that $n \in \text{vis}$ holds. By Definition 6.3, WW totally orders visible nodes writing to the same register, so either $n_2 \xrightarrow{\text{WW}} n_1$ or $n_1 \xrightarrow{\text{WW}} n_2$ holds. Recall that lin is a topological sort of the order including WW and RW. Consequently, it can only be the case that $n_1 \xrightarrow{\text{WW}} n_2$ holds. By Definition 6.3, $n \xrightarrow{\text{RW}} n_2$ holds too, and, hence, $\alpha <_{\text{lin}} \beta_2$. Thus, we have shown that $\beta_1$ is the last write to $x$ that precedes $\alpha$ and that is not located in an aborted or live transaction, meaning that $(\alpha, \alpha')$ is legal. □

### B.3 Proof of Theorem 6.6

**Lemma B.8.** *Let a DRF history $H$ and its graph $G = (N, \_, \text{HB}, \text{WR}, \text{WW}, \text{RW}) \in \text{Graph}(H)$ be such that $(\text{HB} ; (\text{WR} \cup \text{WW} \cup \text{RW}))$ is irreflexive. Then for all nodes $n, n' \in N$, such that at most one of them is a transaction:*

$$n \xrightarrow{\text{WR} \cup \text{WW} \cup \text{RW}} n' \implies n \xrightarrow{\text{HB}} n'.$$

*Proof.* Let us first consider two nodes $n = v$ and $n' = v'$ that are non-transactional accesses such that $v \xrightarrow{\text{WR} \cup \text{WW} \cup \text{RW}} v'$. All non-transactional actions are totally ordered by $\text{cl}(H)$, and, therefore, by $\text{hb}(H)$. Consequently, at least one of the two edges, $v \xrightarrow{\text{HB}} v'$ and $v' \xrightarrow{\text{HB}} v$, is present in the graph $G$. Knowing that $(\text{HB} ; (\text{WR} \cup \text{WW} \cup \text{RW}))$ is irreflexive, we conclude that only the former edge is possible, i.e., $v \xrightarrow{\text{HB}} v'$ holds.

Let us now consider a node-transaction $n = T$ and a non-transactional access $n' = v$ such that $T \xrightarrow{\text{WR} \cup \text{WW} \cup \text{RW}} v$ (the case when $n$ is a non-transactional access and $n'$ is a transaction is analogous). We first assume that $T$ and $v$ are by the same thread. Then either $T \xrightarrow{\text{HB}} v$ or $v \xrightarrow{\text{HB}} T$ holds, since actions of $T$ and $v$ are related by program order accordingly. Knowing that $(\text{HB} ; (\text{WR} \cup \text{WW} \cup \text{RW}))$ is irreflexive, we conclude that it is only possible that $T \xrightarrow{\text{HB}} v$.

We now consider $T$ and $v$ that are by different threads. Let $\alpha'$ be the request action of $v$. It is easy to see from Definition 6.3 of edges of the opacity graph that there exists an action $\alpha \in T$ and a register $x$ such that $\alpha$ and $\alpha'$ access $x$. Since $\alpha$ and $\alpha'$ are by different threads and access the same register, they form a conflict. However, $H$ is DRF, so by Definition 3.2, either $\alpha <_{\text{hb}(H)} \alpha'$ or $\alpha' <_{\text{hb}(H)} \alpha$ holds. When lifted to nodes of the opacity graph, it is the case that either $T \xrightarrow{\text{HB}} v$ or $v \xrightarrow{\text{HB}} T$. Knowing that $(\text{HB} ; (\text{WR} \cup \text{WW} \cup \text{RW}))$ is irreflexive, we conclude that it is only possible that $T \xrightarrow{\text{HB}} v$. □

Given a history $H$ and a graph $G = (H, \_, \_, \text{WR}, \_, \_)$, we write $n \xrightarrow{\text{TXWR}} n'$ when $n, n' \in \text{txns}(H)$ and $n \xrightarrow{\text{WR}} n'$. We also use a shorthand $\text{xpotxwr}(H) = \bigcup_{x \in \text{Reg}}(\text{xpo}(H) ; \text{txwr}_x(H))$.

**Definition B.9.** A hb-path $\pi$ in a history $H$ is a (possibly empty) sequence of triples:

$$\pi = (\alpha_1, R_1, \alpha_2)(\alpha_2, R_2, \alpha_3) \ldots (\alpha_{m-1}, R_{m-1}, \alpha_m),$$

such that:

- $\alpha_1, \alpha_2, \ldots, \alpha_m \in \text{act}(H)$;
- each $R_i$ $(1 \le i \le m-1)$ is one of the following relations: $\text{po}(H), \text{cl}(H), \text{af}(H), \text{bf}(H)$ and $\text{xpotxwr}(H)$.
- for every $i$ such that $1 \le i \le m - 1$, $\alpha_i <_{R_i} \alpha_{i+1}$.

We let $\text{Paths}(H)$ denote the set of all hb-paths in $H$.

**Definition B.10.** Given a history $H$ and transactions $T, T' \in \text{txns}(H)$, we let $\text{TXPaths}_n(H, T, T') \subseteq \text{Paths}(H)$ denote the set of all hb-paths satisfying the following conditions:

- the hb-path starts with an action of $T$ and ends with an action of $T'$;
- $n$ is the number of occurrences of triples of the form $(\_, \text{xpotxwr}(H), \_)$ in the hb-path.

**Lemma B.11.** *For a history $H \in \text{History}$, an opacity graph $G = (N, \text{vis}, \text{HB}, \text{WR}, \text{WW}, \text{RW}) \in \text{Graph}(H)$ and two distinct transactions $T, T' \in \text{txns}(H)$, if $T \xrightarrow{\text{HB}} T'$, then $T \xrightarrow{\text{RT} \cup \text{TXWR}}{}^+ T'$.*

*Proof.* By Definition 6.3, $T \xrightarrow{\text{HB}} T'$ if and only if there exist $\alpha \in T$ and $\alpha' \in T'$ such that $\alpha <_{\text{hb}(H)} \alpha'$. It is easy to see that $T \xrightarrow{\text{HB}} T'$ holds if and only if $\bigcup_{n \ge 0} \text{TXPaths}_n(H, T, T') \ne \emptyset$. Thus, to prove the lemma, we demonstrate that for a given history $H$ and its graph $G = (N, \text{vis}, \text{HB}, \text{WR}, \text{WW}, \text{RW}) \in \text{Graph}(H)$, the following holds:

$$\forall T, T' \in \text{txns}(H). T \ne T' \wedge \bigcup_{n \ge 0} \text{TXPaths}_n(H, T, T') \ne \emptyset \implies$$

$$T \xrightarrow{\text{RT} \cup \text{TXWR}}{}^+ T'. \quad (3)$$

We define $\Phi(n)$ as the following auxiliary statement:

$$\forall T, T' \in \mathsf{txns}(H).\, T \neq T' \wedge \mathsf{TXPaths}_n(H, T, T') \neq \emptyset \implies$$
$$T \xrightarrow{\mathsf{RT} \cup \mathsf{TXWR}}{}^+ T'.$$

To prove (3), we show that $\Phi(n)$ holds for all $n \geq 0$ by induction on $n$.

BASE OF THE INDUCTION. Let us consider two distinct transactions $T$ and $T'$ and the hb-paths $\mathsf{TXPaths}_0(H, T, T')$ between them. Since these hb-paths begin and end in different transactions and only feature relations $\mathsf{po}(H)$, $\mathsf{cl}(H)$, $\mathsf{af}(H)$ and $\mathsf{bf}(H)$, every hb-path $\mathsf{TXPaths}_0(H, T, T')$ includes actions denoting the end of $T$ and the beginning of $T'$. Hence, there exists a path $\pi = (\alpha, \_, \_) \ldots (\_, \_, \alpha') \in \mathsf{TXPaths}_0(H, T, T')$ such that $\alpha$ is the end of a transaction $T$ and $\alpha'$ is the beginning of a transaction $T'$. Since relations $\mathsf{po}(H)$, $\mathsf{cl}(H)$, $\mathsf{af}(H)$ and $\mathsf{bf}(H)$ are all consistent with the history execution order $<_H$ (see §3), $\alpha <_H \alpha'$. Therefore, $T \xrightarrow{\mathsf{RT}} T'$ holds.

INDUCTION STEP. Assuming that $\Phi(i)$ holds for $i \leq n$, we demonstrate that so does $\Phi(n+1)$. Let us consider two distinct transactions $T, T' \in \mathsf{txns}(H)$ and a hb-path $\pi \in \mathsf{TXPaths}_{n+1}(H, T, T')$.

Consider actions $\alpha_1$ and $\alpha_2$ such that the triple $(\alpha_1, \mathsf{xpotxwr}(H), \alpha_2)$ occurs the latest in $\pi$: there exist $\pi'$ and $\pi''$ such that $\pi = \pi'(\alpha_1, \mathsf{xpotxwr}(H), \alpha_2)\pi''$ and there is no triple $(\_, \mathsf{xpotxwr}(H), \_)$ in $\pi''$. By definition of $\mathsf{xpo}(H)$ and $\mathsf{txwr}(H)$, there exists a transaction $T_1$ and actions $\alpha_{\mathsf{begin}}$ and $\alpha_{\mathsf{write}}$ such that:

- $\alpha_{\mathsf{begin}}$ denotes a beginning of a transaction $T_1$;
- $\alpha_{\mathsf{write}}$ denotes a write request action by $T_1$;
- $\alpha_1 <_{\mathsf{po}(H)} \alpha_{\mathsf{begin}} <_{\mathsf{po}(H)} \alpha_{\mathsf{write}} <_{\mathsf{txwr}\_(H)} \alpha_2$ holds.

To conclude the induction step, we make the following three observations. Firstly, note that the hb-path $\pi'(\alpha_1, \mathsf{po}(H), \alpha_{\mathsf{write}}) \in \mathsf{TXPaths}_n(H, T, T_1)$, and, therefore, $T \xrightarrow{\mathsf{RT} \cup \mathsf{TXWR}}{}^+ T_1$ holds by the induction hypothesis $\Phi(n)$. Secondly, let $T_2$ be the transaction of $\alpha_2$. Since $\alpha_{\mathsf{write}} <_{\mathsf{txwr}\_(H)} \alpha_2$ holds, so does $T_1 \xrightarrow{\mathsf{TXWR}} T_2$. Finally, $\pi'' \in \mathsf{TXPaths}_0(H, T_2, T')$. By the induction hypothesis $\Phi(0)$, $T_2 \xrightarrow{\mathsf{RT} \cup \mathsf{TXWR}}{}^+ T'$ holds. Altogether, the three observations imply $T \xrightarrow{\mathsf{RT} \cup \mathsf{TXWR}}{}^+ T'$. □

*Proof of Theorem 6.6.* Let us consider a DRF history $H$ and its graph $G = (N, \mathsf{vis}, \mathsf{HB}, \mathsf{WR}, \mathsf{WW}, \mathsf{RW}) \in \mathsf{Graph}(H)$. Consider a cycle $\pi$ in $G$. Without loss of generality, we can assume that all vertices on the cycle are distinct.

We first consider the case when all nodes in the cycle $\pi$ are non-transactional accesses. Note that non-transactional accesses of $H$ are totally ordered by $\mathsf{cl}(H)$. Knowing that $\mathsf{cl}(H)$ is consistent with the execution order $<_H$ of the history $H$, we conclude that there cannot be such cycle $\pi$.

In order to construct a cycle $\pi'$ over transactions only with edges from $\mathsf{RT} \cup \mathsf{DEP}$, we consider any two adjacent transactions in the cycle $\pi$, i.e., such transactions $T$ and $T'$ that no other transaction appears between $T$ and $T'$. We then demonstrate that in the opacity graph $G$, there is a path from $T$ to $T'$ in $\mathsf{RT} \cup \mathsf{DEP}$ over transactions only, and we add the path to $\pi'$.

There are two possibilities: either $T'$ immediately follows $T$ in the cycle, or they are separated by non-transactional accesses.

When $T'$ immediately follows $T$ in the cycle $\pi$, the two transactions are connected by an opacity graph edge $\mathsf{HB} \cup \mathsf{DEP}$. If $T \xrightarrow{\mathsf{HB}} T'$, then by Lemma B.11, $T \xrightarrow{\mathsf{RT} \cup \mathsf{TXWR}}{}^+ T'$, so we add the latter edges to $\pi'$. If $T \xrightarrow{\mathsf{DEP}} T'$, then we add the same edge to $\pi'$.

When $T$ and $T'$ are separated by other actions in the cycle $\pi$, we consider a path between the two transactions:

$$T \xrightarrow{\mathsf{HB} \cup \mathsf{DEP}} n_1 \xrightarrow{\mathsf{HB} \cup \mathsf{DEP}} n_2 \xrightarrow{\mathsf{HB} \cup \mathsf{DEP}} \ldots$$
$$\ldots \xrightarrow{\mathsf{HB} \cup \mathsf{DEP}} n_K \xrightarrow{\mathsf{HB} \cup \mathsf{DEP}} T', \quad (4)$$

where each $n_i \in \mathsf{nontxn}(H)$ ($i = 1..K$) is a node of the graph $G$ denoting a non-transactional access. By Lemma B.8, all non-HB edges on the path can be replaced by HB:

$$T \xrightarrow{\mathsf{HB}} n_1 \xrightarrow{\mathsf{HB}} n_2 \xrightarrow{\mathsf{HB}} \ldots \xrightarrow{\mathsf{HB}} n_K \xrightarrow{\mathsf{HB}} T'$$

Therefore, $T \xrightarrow{\mathsf{HB}} T'$ holds. By Lemma B.11, there is a sequence of edges $T \xrightarrow{\mathsf{RT} \cup \mathsf{TXWR}}{}^+ T'$. We add all of those edges to the cycle $\pi'$. □

Analogously to Theorem 6.6, we can prove the following lemma. Let txWR, txWW, txRW denote WR, WW and RW dependencies between transactions accordingly.

**Lemma B.12.** *Let a data-race free history $H$ and an opacity graph $G = (N, \mathsf{vis}, \mathsf{HB}, \mathsf{WR}, \mathsf{WW}, \mathsf{RW}) \in \mathsf{Graph}(H)$ be such that the relation $(\mathsf{HB} \,;\, (\mathsf{WR} \cup \mathsf{WW} \cup \mathsf{RW}))$ is irreflexive. If $T \xrightarrow{\mathsf{HB} \cup \mathsf{WR} \cup \mathsf{WW} \cup \mathsf{RW}}{}^+ T'$ holds, then so does $T \xrightarrow{\mathsf{RT} \cup \mathsf{txWR} \cup \mathsf{txWW} \cup \mathsf{txRW}}{}^+ T'.$*

That is, if there exists a path over edges $\mathsf{HB} \cup \mathsf{WR} \cup \mathsf{WW} \cup \mathsf{RW}$ between two distinct transactions $T$ and $T'$ of the graph $G$, there also exists a path consisting only of edges from $\mathsf{RT} \cup \mathsf{txWR} \cup \mathsf{txWW} \cup \mathsf{txRW}$ (note that RT edges are not necessarily present in the graph $G$).

```
1  Value clock, reg[NRegs], ver[NRegs];
2  Lock lock[NRegs];
3  Bool active[NThreads];
4  Set<Register> rset; // for each transaction
5  Map<Register, Value> wset; // for each transaction
6  Value rver; // for each transaction, initially ⊥
7  Value wver; // for each transaction, initially ⊤
8
9  function txbegin(Transaction T):
10   active[threadOf(T)] := true;
11   rver[T] := clock;
12   return;
13
14 function read(Transaction T, Register x):
15   if (wset[T].contains(x))
16     return wset[T].get(x);
17   ts1 := ver[x];
18   value := reg[x];
19   locked := lock[x].test();
20   ts2 := ver[x];
21   if (locked ∨ ts1 ≠ ts2 ∨ rver[T] < ts2)
22     return abort(T);
23   rset[T].put(x);
24   return value;
25
26 function write(Transaction T, Register x, Value v):
27   wset[T].put(x, v);
28   return;
29
30 function fence():
31   Bool r[NThreads]; // initially all false
32   foreach t in ThreadID:
33     r[t] := active[t];
34   foreach t in ThreadID:
35     if (r[t]):
36       while (active[t]);
37   return;

30 function txcommit(Transaction T):
31   Set<Lock> lset := ∅;
32   foreach x in wset[T]:
33     Bool locked := lock[x].trylock();
34     if (¬locked):
35       lset.add(x);
36     else:
37       foreach y in lset[T]:
38         lock[y].unlock();
39       return abort(T);
40   wver[T] := fetch_and_increment(clock)+1;
41   foreach x in rset[T]:
42     Bool locked := lock[x].test();
43     atomic {
44       Value ts := ver[x];
45       pv[T][x] := ¬(locked ∨ rver[T] < ts);
46     }
47     if (locked ∨ rver[T] < ts):
48       foreach y in lset[T]:
49         lock[y].unlock();
50       return abort(T);
51   foreach (x, v) in wset[T]:
52     reg[x] := v;
53     ver[x] := wver[T];
54     lock[x].unlock();
55   return commit(T);
56
57 function abort(Transaction T):
58     return aborted;
59     active[threadOf(T)] := false;
60
61 function commit(Transaction T):
62     return committed;
63     active[threadOf(T)] := false;
```

**Figure 9.** TL2 pseudocode.

## C Strong opacity of TL2

In this section, we provide details for the proof of strong opacity of TL2. To this end, we first argue that histories of TL2 are well-formed, and in the rest of the section we discharge the proof obligations arising from graph characterization of TL2 histories by means of Lemma 6.4 and Theorem 6.6.

### C.1 Preliminaries

In Figure 9 we present the full pseudo-code of the TL2 software TM implementation. In the code, we use NThreads to denote the number of threads and threadOf(T) to denote the thread executing a given transaction T. We also let ⊥ and ⊤ denote special minimal and maximal values, which rver[T] and wver[T] are initialized to (for every transaction T). The value ⊥ is also used to describe a state of a lock as follows. We assume that the lock lock[x] of each register x is either unlocked or stores an identifier of a transaction holding the lock, i.e., Lock = {⊥} ⊎ Transaction. Thus, when lock[x] = ⊥, the lock is not acquired by any transaction, and when lock[x] = T, we know that T is holding a lock on x.

Functions txbegin, txcommit, write, read and fence of the pseudocode generate corresponding request and response actions from Figure 4 simultaneously with an invocation and return from the function accordingly. Additionally, we assume that each transaction $T$ upon aborting or committing executes a handler aborted($T$) or committed($T$) accordingly, both of which simply unset the active[$t$] flag (at line 59 or line 63) after appending an abort-response or a commit-response.

In the proof of strong opacity of TL2, we consider every execution of the most general client of TL2. At each step of it, we maintain a triple $(s, H, G)$ consisting of a current state $s$, a current history $H$ and its matching opacity graph $G \in \text{Graph}(H)$. In addition to these three components, for every transaction $T$ we also maintain a *ghost* variable pv[$T$] : Register $\rightarrow$ Bool that maps a register to true only if $T$ post-validated a read from it. Ghost variables are different from usual variables in that they do not describe the concrete state of the execution, and are means to representing information about the past of executions in proofs. For simplicity of presentation, in the following we treat the pv variables as if they were a part of concrete state.

## C.2 Well-formedness of TL2 histories

We argue that the histories of TL2 are well-formed. The most non-trivial well-formedness property of histories is that of the fences, i.e., we need to show that:

- Fence blocks until all active transactions complete: if $\tau = \tau_1 \, (\_, t, \text{txbegin}) \, \tau_2 \, (\_, t', \text{fbegin}) \, \tau_3 \, (\_, t', \text{fend}) \, \tau_4$ then either $\tau_2$ or $\tau_3$ contains an action of the form $(\_, t, \text{committed})$ or $(\_, t, \text{aborted})$.

Let us consider the fence implementation at lines 30–36 of Figure 9. The fence function consists of two loops: first, it iterates over all threads and records whether each thread $t$ has an active transaction in a local variable r[$t$]; and then for each thread $t$, the fence waits until active[$t$] becomes false, if r[$t$] is true.

Let us consider execution of a fence in a thread $t'$ and let $A$ be the set of all active transactions at the beginning of the execution of the fence. For each transaction $T \in A$ in a thread $t$, there are two possibilities:

- The transaction $T$ commits (aborts) before the fence checks active[$t$] for the first time. Therefore, its commit-response $(\_, t, \text{committed})$ (abort-response $(\_, t, \text{aborted})$) occurs before $(\_, t', \text{fend})$ in a history of the execution.
- The transaction $T$ commits (aborts) after the fence checks active[$t$] for the first time, meaning that active[$t$] = true at that point. When that is the case, the fence sets r[$t$] to true. Later on at line 36, the fence repeatedly re-reads the value of active[$t$] until it observes false. Note that $T$ will append a commit-response $(\_, t, \text{committed})$ (abort-response $(\_, t, \text{aborted})$) to the history before setting active[$t$] to false. Therefore, by the time the fence reads false from active[$t$], the $T$'s commit-response $(\_, t, \text{committed})$ (abort-response $(\_, t, \text{aborted})$) is already in the history, meaning that it occurs before $(\_, t', \text{fend})$.

Based on these observations, we conclude that fences block until all active transactions complete, and that histories of TL2 are well-formed.

## C.3 Opacity graph construction

The construction of the graph is inductive in the length of the execution of the most general client. At each step of the construction, we maintain a triple $(s, H, G)$ consisting of the concrete state $s$, a history $H$ and its opacity graph $G \in \text{Graph}(H)$. We start from the initial state[3] empty history and an empty graph, and modify them as the execution proceeds. To this end, simultaneously with certain primitive commands of the TL2 algorithm in Figure 9 we execute *graph updates* modifying the opacity graph.

In Figure 10, we specify all the changes each graph update performs to an opacity graph $G = (N, \text{vis}, \text{HB}, \text{WR}, \text{WW}, \text{RW})$ of the triple $(s, H, G)$. The updates make use of auxiliary predicates reads($n, x, v$) and writes($n, x, v$), which we further define:

- A predicate reads($n, x, v$) holds of $(s, H, G)$, if either $n \in N$ reads $v$ from $x$ non-transactionally or $n \in N$ is a transaction containing a non-local read of $v$ from $x$.
- A predicate writes($n, x, v$) holds of $(s, H, G)$, if $n$ is a node of the graph $G$ such that either $n = ((\_, \_, \text{write}(x, v)), (\_, \_, \text{ret}(\bot))) \in \text{nontxn}(H)$ holds, or $n \in \text{txns}(H)$ and its last write to $x$ is $((\_, \_, \text{write}(x, v)), (\_, \_, \text{ret}(\bot)))$.

We perform changes instructed by graph updates simultaneously with certain transitions of the TL2 algorithm, which we further specify for each update:

- TXBEGIN($T$) occurs at the beginning of a transaction $T$;
- TXREAD($T, x, v$) occurs when a transaction $T$ returns a value $v$ at line 24, and a corresponding read response is appended to the current history;
- TXVIS($T$) occurs simultaneously with the last transition of the loop at lines 41–50 in the transaction $T$;
- NTXREAD($\nu, x, v$) occurs when a non-transactional read access $\nu$ reads a value $v$ from a register $x$;
- NTXWRITE($\nu, x$) occurs when a non-transactional write access $\nu$ writes a value to a register $x$.

## C.4 Invariants

We prove strong opacity of TL2 by demonstrating that at each step of the construction of an opacity graph characterized by $(s, H, G)$, where $s$ is a concrete state, $H$ is a history and $G$ is its matching opacity graph, the triple satisfies a global

---

[3] the state, in which every registers stores the initial value $v_{\text{init}}$ and has a version 0; clock stores 0; no lock is held and no transaction is active in any thread.

TXBEGIN($T$):
  $N := N \cup \{T\}$;
  vis($T$) := false;
  HB := HB $\cup \{n \xrightarrow{\text{HB}} T \mid \exists \alpha \in n, \alpha' \in T.\ \alpha <_{\text{hb}(H)} \alpha'\}$;

TXREAD($T, x, v$):
  **if** ($v = v_{\text{init}}$):
    RW := RW $\cup \{T \xrightarrow{\text{RW}_x} n \mid \text{vis}(n) \wedge \text{writes}(n, x, \_)\}$
  **else**:
    WR := WR $\cup \{n \xrightarrow{\text{WR}_x} T \mid \text{writes}(n, x, v)\}$;
    RW := RW $\cup \{T \xrightarrow{\text{RW}_x} n' \mid \text{writes}(n, x, v) \wedge n \xrightarrow{\text{WW}_x} n'\}$;
    HB := HB $\cup \{n \xrightarrow{\text{HB}} T \mid \exists T', n'.\ \text{writes}(T', x, v)$
           $\wedge \text{threadOf}(T') = \text{threadOf}(n')$
           $\wedge n \xrightarrow{\text{HB}}^* n' \xrightarrow{\text{HB}} T'\}$;

TXVIS($T$):
  vis($T$) := true;
  **foreach** $x$ **in** wset[$T$]:
    WW := WW $\cup \{n \xrightarrow{\text{WW}_x} T \mid n \neq T \wedge \text{vis}(n)$
                   $\wedge \text{writes}(n, x, \_)\}$;
    RW := RW $\cup \{n \xrightarrow{\text{RW}_x} T \mid n \neq T \wedge \text{reads}(n, x, \_)\}$;

NTXREAD($v, x, v$):
  $N := N \cup \{v\}$;
  vis($v$) := true;
  **if** ($v \neq v_{\text{init}}$):
    WR := WR $\cup \{n \xrightarrow{\text{WR}_x} T \mid \text{writes}(n, x, v)\}$;
    HB := HB $\cup \{n \xrightarrow{\text{HB}} T \mid \exists \alpha \in n, \alpha' \in v.\ \alpha <_{\text{hb}(H)} \alpha'\}$;

NTXWRITE($v, x$):
  $N := N \cup \{v\}$;
  vis($v$) := true;
  WW := WW $\cup \{n \xrightarrow{\text{WW}_x} v \mid n \neq v \wedge \text{vis}(n) \wedge \text{writes}(n, x, \_)\}$;
  RW := RW $\cup \{n \xrightarrow{\text{RW}_x} v \mid n \neq v \wedge \text{reads}(n, x, \_)\}$;
  HB := HB $\cup \{n \xrightarrow{\text{HB}} v \mid \exists \alpha \in n, \alpha' \in v.\ \alpha <_{\text{hb}(H)} \alpha'\}$;

**Figure 10.** Description of graph updates

invariant presented in Figure 11. The invariant makes use of the following auxiliary definitions:

- Given a relation $R$ on a set of graph nodes and a node $n$, we say that isLastIn($R, n$) holds, if $R$ is a linear order and $n$ is the last node in $R$.
- We let completed($T$) be a predicate that holds of $(s, H, G)$, if $T \in \text{txns}(H)$ is a committed or aborted transaction. We also introduce a predicate aborted($T$) that holds of $(s, H, G)$, when $T \in \text{txns}(H)$ is an aborted transaction.

The most important invariant properties correspond to the proof obligations arising from Lemma 6.4 and Theorem 6.6. Recall that we need to demonstrate that for a history $H \in \mathcal{H}_{\mathbb{C}}|_{\text{DRF}}$ and the graph $G \in \text{Graph}(H)$ that we construct, the following holds:

1. $H$ is consistent;
2. if $G = (N, \text{vis}, \text{HB}, \text{WR}, \text{WW}, \text{RW})$, then the relation (HB ; (WR $\cup$ WW $\cup$ RW)) is irreflexive;
3. $G$ does not contain a cycle over transactions only with edges from RT $\cup$ WR $\cup$ WW $\cup$ RW.

The obligations 1–3 are discharged by proving invariants INV.2, INV.3 and INV.4 accordingly. In the following, we explain each part of the invariant.

The invariant INV.1 requires that histories be data-race free, since we only need to consider histories from $\mathcal{H}_{\mathbb{C}}|_{\text{DRF}}$ in the proof. The invariant INV.2(a) asserts consistency of TL2 histories, while INV.2(b) is an auxiliary property relating a predicate writes($\_, \_, \_$) with the content of write-sets of transactions. The invariant INV.3 requires that (HB ; (WR $\cup$ WW $\cup$ RW)) be irreflexive. The invariant INV.4 asserts that the opacity graph $G$ does not contain cycles over transactions in (RT $\cup$ txWR $\cup$ txWW $\cup$ txRW). By Theorem 6.6, together the latter two invariants imply that the graph $G$ is acyclic.

The invariants mentioned so far are not inductive, so we strengthen them with additional auxiliary invariants. To this end, INV.5 relates the order of timestamps to the edges of the opacity graph. In order to maintain this invariant inductively, we include an additional invariant INV.6, which helps to establish that INV.5(c–e) are preserved by the graph update TXVIS($\_$). To this end, it asserts the properties of every commit-pending transaction $T'$ that has acquired a lock on a register, but has not added corresponding write and anti-dependencies yet.

The invariant INV.7 asserts various well-formedness properties of read and write timestamps of each transaction, and the invariant INV.8 asserts multiple well-formedness conditions relating the concrete state, history actions and graph edges.

**Lemma C.1.** *The invariant* INV *is preserved by the graph updates* TXREAD($T, x$), TXVIS($T$), NTXREAD($v$), NTXWRITE($v$) *and* TXBEGIN($T$).

The invariant INV.1 does not require a proof of preservation. Indeed, in the Fundamental Property (Theorem 5.3) data-race freedom is an obligation that clients of a TM system need to fulfill, not a TM implementation. Hence, in the proof of strong opacity of TL2, INV.1 is an assumption. Also, it is easy to see that the well-formedness conditions of INV.7 and INV.8 are preserved trivially by construction of histories and graphs. We provide proof details for the remaining invariants in the following sections:

INV denotes the smallest set of triples $(s, H, G)$ of a concrete state $s$, a history $H$ and an opacity graph $G = (N, \text{vis}, \text{HB}, \text{WR}, \text{WW}, \text{RW}) \in \text{Graph}(H)$, all satisfying the following:

1. The history $H$ is data-race free.
2. Consistency invariants:
   a. $H$ is a consistent history.
   b. The write-set of each transaction $T$ consists of its most recent writes to registers:
   $$\forall T \in \text{txns}(H), x \in \text{Reg}, v \in \mathbb{Z}.\ (x, v) \in \text{wset}[T] \iff \text{writes}(T, x, v)$$
3. The relation $(\text{HB} ; (\text{WR} \cup \text{WW} \cup \text{RW}))$ is irreflexive.
4. The relation $(\text{RT} \cup \text{txWR} \cup \text{txWW} \cup \text{txRW})$ is acyclic.
5. Read and write timestamps of transactions have the following properties:
   a. $\forall T, T' \in \text{txns}(H).\ T \xrightarrow{\text{RT}} T' \implies ((\text{vis}(T) \implies \text{wver}[T] \leq \text{rver}[T']) \land (\neg\text{vis}(T) \implies \text{rver}[T] \leq \text{rver}[T']))$
   $\lor\ \text{rver}[T'] = \bot$
   b. $\forall T, T' \in \text{txns}(H).\ T \xrightarrow{\text{WR}} T' \implies \text{wver}[T] \leq \text{rver}[T']$
   c. $\forall T, T' \in \text{txns}(H).\ T \xrightarrow{\text{WW}} T' \implies \text{wver}[T] < \text{wver}[T']$
   d. $\forall T, T' \in \text{txns}(H).\ T \xrightarrow{\text{RW}} T' \implies \text{rver}[T] < \text{wver}[T']$
   e. $\forall T, T' \in \text{txns}(H).\ T \xrightarrow{\text{RW}_x} T' \land \text{pv}[T][x] = \text{true} \implies \text{wver}[T] < \text{wver}[T']$
6. Commit-pending transaction $T'$ has the following properties:
   a. $\forall T, T' \in \text{txns}(H), x \in \text{Reg}.\ T \neq T' \land \text{writes}(T, x) \land \text{vis}(T) \land \text{lock}[x] = T' \implies \text{wver}[T] < \text{wver}[T']$
   b. $\forall T, T' \in \text{txns}(H), x \in \text{Reg}.\ T \neq T' \land \text{reads}(T, x, \_) \land \text{lock}[x] = T' \implies \text{rver}[T] < \text{wver}[T']$
   c. $\forall T, T' \in \text{txns}(H), x \in \text{Reg}.\ T \neq T' \land \text{pv}[T][x] = \text{true} \land \text{lock}[x] = T' \implies \text{wver}[T] < \text{wver}[T']$
7. Well-formedness properties of read and write timestamps:
   a. $\forall T \in \text{txns}(H).\ \text{rver}[T] < \text{wver}[T]$
   b. $\forall T \in \text{txns}(H).\ \text{rver}[T] \leq \text{clock} \land (\text{wver}[T] = \top \lor \text{wver}[T] \leq \text{clock})$
   c. $\forall T \in \text{txns}(H).\ \text{wver}[T] \neq \top \implies \text{rver}[T] \neq \bot$
   d. $\forall T \in \text{txns}(H).\ \text{reads}(T, \_, \_) \implies \text{rver}[T] \neq \bot)$
   e. $\forall T \in \text{txns}(H).\ (\exists x.\ \text{pv}[T][x] = \text{true}) \lor \text{vis}(T) \implies \text{wver}[T] \neq \top$
8. Auxiliary invariants:
   a. The value of each unlocked register $x$ is ether the initial $v_{\text{init}}$ or the value written by the last node in $\text{WW}_x$.
   $$\forall x \in \text{Reg}.\ \text{lock}[x] = \bot \implies (\text{reg}[x] = v_{\text{init}} \iff \neg\exists n.\ \text{vis}(n) \land \text{writes}(n, x, \_))$$
   $$\land\ (\text{reg}[x] \neq v_{\text{init}} \iff \exists n.\ \text{isLastIn}(\text{WW}_x, n) \land \text{writes}(n, x, \text{reg}[x]))$$
   b. The version of each unlocked register $x$ is either the initial version $v_{\text{init}}$ or the write timestamp of the last transaction in $\text{txWW}_x$.
   $$\forall x \in \text{Reg}.\ \text{lock}[x] = \bot \implies (\text{ver}[x] = v_{\text{init}} \iff \neg\exists T \in \text{txns}(H).\ \text{vis}(T) \land \text{writes}(T, x, \_))$$
   $$\land\ (\text{ver}[x] \neq v_{\text{init}} \iff \exists T \in \text{txns}(H).\ \text{isLastIn}(\text{txWW}_x, T) \land \text{ver}[x] = \text{wver}[T])$$
   c. Visible transactions have their reads post-validated:
   $$\forall T \in \text{txns}(H), x \in \text{Reg}.\ \text{vis}(T) \land \text{reads}(T, x, \_) \implies \text{pv}[T][x] = \text{true}$$
   d. HB-edges do not originate from active transactions:
   $$\forall T \in \text{txns}(H).\ \neg\text{completed}(T) \implies \neg\exists n'.\ T \xrightarrow{\text{HB}} n'$$
   e. A transaction holding a lock on a register is not completed and writes to that register and is not overwritten:
   $$\forall x \in \text{Reg}, T \in \text{txns}(H).\ \text{lock}[x] = T \implies \neg\text{completed}(T) \land \text{writes}(T, x, \_) \land \neg\exists T'.\ T \xrightarrow{\text{WW}_x} T'$$

**Figure 11.** The TL2 invariant

| Section # | Invariant |
|---|---|
| Section C.6 | INV.2 |
| Section C.7 | INV.3 |
| Section C.8 | INV.5 |
| Section C.9 | INV.4 |
| Section C.10 | INV.6 |

Note that we are considering the invariants out of the ascending order due to Section C.9 depending on Section C.8.

## C.5 Timestamp order properties

We state Proposition C.2, which relates paths between transactions with timestamp order, and then we use it in multiple proofs of preservation of the invariant.

**Proposition C.2.** *If $(s, H, G)$ satisfies INV.5, INV.7 and INV.8, then for every transactions $T, T' \in \text{txns}(H)$, if $T \xrightarrow{\text{RT} \cup \text{txWR} \cup \text{txWW} \cup \text{txRW}}{}^+ T'$, then either of the following holds:*

1. $\text{vis}(T) \wedge \text{vis}(T') \implies \text{wver}[T] < \text{wver}[T']$;
2. $\neg\text{vis}(T) \wedge \text{vis}(T') \implies \text{rver}[T] < \text{wver}[T']$;
3. $\text{vis}(T) \wedge \neg\text{vis}(T') \implies \text{wver}[T] \le \text{rver}[T'] \vee \text{rver}[T'] = \bot$;
4. $\neg\text{vis}(T) \wedge \neg\text{vis}(T') \implies \text{rver}[T] \le \text{rver}[T'] \vee \text{rver}[T'] = \bot$.

*Proof of Proposition C.2.* The proof is by induction on the length of the path $T \xrightarrow{\text{RT} \cup \text{txWR} \cup \text{txWW} \cup \text{txRW}}{}^+ T'$. Let $\phi_1(T, T')$, $\phi_2(T, T')$, $\phi_3(T, T')$ and $\phi_4(T, T')$ be predicates corresponding to the implications of the proposition. For each $k \ge 1$, we prove $\Phi(k)$, which is defined as follows:

$$\Phi(k) \triangleq \forall T, T' \in \text{txns}(H). \, T \xrightarrow{\text{RT} \cup \text{txWR} \cup \text{txWW} \cup \text{txRW}}{}^k T'$$
$$\implies \phi_1(T, T') \wedge \phi_2(T, T') \wedge \phi_3(T, T') \wedge \phi_4(T, T')$$

**Base of induction.** We need to show that $\Phi(1)$ holds. Let us consider any two transactions $T$ and $T'$ in the opacity graph and assume that $T \xrightarrow{\text{RT} \cup \text{WR} \cup \text{WW} \cup \text{RW}} T'$ holds. We consider each possible edge separately and demonstrate $\phi_1(T, T')$, $\phi_2(T, T')$, $\phi_3(T, T')$ or $\phi_4(T, T')$ depending on visibility of $T$ and $T'$.

1. Consider the edge $T \xrightarrow{\text{RT}} T'$. According to the invariant INV.5(a), we consider make the following conclusions depending on visibility of $T$ and $T'$.

   - Let us assume that both $T$ and $T'$ are visible. We demonstrate that $\phi_1(T, T')$ holds. From the invariant INV.5(a), we know that either $\text{wver}[T] \le \text{rver}[T']$ or $\text{rver}[T'] = \bot$ is the case. Note that the contra-positive of the invariant INV.7(c,e) states that $\text{rver}[T'] = \bot$ contradicts visibility of $T'$. Therefore, only $\text{wver}[T] \le \text{rver}[T']$ can hold in this case. By INV.7(a), $\text{rver}[T'] < \text{wver}[T']$ holds, which allows us to conclude $\text{wver}[T] \le \text{wver}[T']$ (coinciding with $\phi_1(T, T')$).

   - Let us assume that $T$ is not visible and $T'$ is. We demonstrate that $\phi_2(T, T')$ holds. From the invariant INV.5(a), we know that either $\text{rver}[T] \le \text{rver}[T']$ or $\text{rver}[T'] = \bot$ is the case. Note that the contra-positive of the invariant INV.7(c,e) states that $\text{rver}[T'] = \bot$ contradicts visibility of $T'$. Therefore, only $\text{rver}[T] \le \text{rver}[T']$ can hold in this case. By INV.7(a), $\text{rver}[T'] < \text{wver}[T']$ holds, which allows us to conclude $\text{rver}[T] < \text{wver}[T']$ (coinciding with $\phi_2(T, T')$).

   - Let us assume that $T$ is visible and $T'$ is not. It is easy to see that $\phi_3(T, T')$ holds, since, by INV.5(a), either $\text{wver}[T] \le \text{rver}[T']$ or $\text{rver}[T'] = \bot$ is true.

   - Let us assume that neither $T$ nor $T'$ is visible. It is easy to see that $\phi_4(T, T')$ holds, since, by INV.5(a), either $\text{rver}[T] \le \text{rver}[T']$ or $\text{rver}[T'] = \bot$ is true.

2. $T \xrightarrow{\text{WR}} T'$. By Definition 6.3, $\text{vis}(T)$ holds. From invariant INV.5(b), we obtain that $\text{wver}[T] \le \text{rver}[T']$ holds. Hence, if $\text{vis}(T')$ does not hold, we can conclude $\phi_3(T, T')$. Let us consider the case when $\text{vis}(T')$ holds. By INV.7(a), $\text{rver}[T'] < \text{wver}[T']$ holds. Therefore, $\text{wver}[T] < \text{wver}[T']$ holds, which allows us to conclude $\phi_1(T, T')$.

3. $T \xrightarrow{\text{WW}} T'$. By Definition 6.3, $\text{vis}(T)$ and $\text{vis}(T')$ both hold. By invariant INV.5(c), $\text{wver}[T] \le \text{wver}[T']$ holds, which allows us to conclude $\phi_1(T, T')$.

4. $T \xrightarrow{\text{RW}} T'$. By Definition 6.3, $\text{vis}(T')$ holds. Let us first assume that $\text{vis}(T)$ does not hold. By INV.5(d), $\text{rver}[T] < \text{wver}[T']$ holds, which allows us to conclude $\phi_2(T, T')$. Let us now consider the case when $\text{vis}(T)$ holds. By INV.8(c), $T$ post-validated all of its reads. Therefore, by INV.5(e), $\text{wver}[T] < \text{wver}[T']$ holds, which allows us to conclude $\phi_1(T, T')$

**Induction step.** Let us consider any $k \ge 1$ and assume that $\Phi(k')$ holds for all $k' \le k$, i.e.:

$$\forall T, T'. \, T \xrightarrow{\text{RT} \cup \text{txWR} \cup \text{txWW} \cup \text{txRW}}{}^{k'} T'$$
$$\implies \phi_1(T, T') \wedge \phi_2(T, T') \quad (5)$$

We need to show that $\Phi(k + 1)$ holds too. To this end, we consider any path $T \xrightarrow{\text{RT} \cup \text{txWR} \cup \text{txWW} \cup \text{txRW}}{}^{k+1} T'$.

Let $T''$ be the next transaction after $T$ on the path, i.e., such that $T \xrightarrow{\text{RT} \cup \text{txWR} \cup \text{txWW} \cup \text{txRW}} T'' \xrightarrow{\text{RT} \cup \text{txWR} \cup \text{txWW} \cup \text{txRW}}{}^k T'$. It is easy to see that $\Phi(k + 1)$ can be obtained from the induction hypotheses $\Phi(1)$ (instantiated for $T$ and $T''$) and $\Phi(k)$ (instantiated for $T''$ and $T'$). □

### C.5.1 Anti-dependencies of read updates

We now formulate and establish a property of anti-dependency edges added at the graph update $\text{TXREAD}(T, x, v)$, which we use in proofs of preservation of several invariants. To this end, we introduce an auxiliary predicate $\text{PAD}(T, x, v)$ denoting a set of triples $(s'', H'', G'')$ such that:

- in the history $H''$, $T$ is a transaction whose last action is a read request $(\_, \_, \text{read}(x))$;
- if $v = v_{\text{init}}$ holds, then

$$\forall n'. \text{writes}(n', x, \_) \land \neg \text{aborted}(n') \implies$$
$$(n' \in \text{txns}(H)) \land \text{rver}[T] < \text{wver}[n']$$

- if $v \neq v_{\text{init}}$ holds, then there exists a node $n$ such that $\text{vis}(n)$, $\text{writes}(n, x, v)$ and the following all hold:

$$\forall n'. \text{writes}(n', x, \_) \land \neg \text{aborted}(n') \land \neg n' \xrightarrow{\text{WW}_x} n \implies$$
$$(n' \in \text{txns}(H)) \land \text{rver}[T] < \text{wver}[n']$$

The following proposition asserts that $\text{PAD}(T, x, v)$ is *stable under interference* from other threads, i.e., transitions and graph updates of other threads do not invalidate it.

**Proposition C.3.** *If $(s, H, G) \in \text{PAD}(T, x, v)$ and $(s', H', G')$ is a result of a graph update or a transition in a thread different from $\text{threadOf}(T)$, then $(s', H', G') \in \text{PAD}(T, x, v)$.*

*Proof.* Note that opacity graphs are constructed monotonously, i.e., once we add an edge, it stays in the graph. Hence, in a proof of stability of $\text{PAD}(T, x, v)$ it is important to consider only graph updates $\text{NTXWRITE}(n', x)$ and $\text{TXVIS}(n')$, since they introduce new nodes writing to $x$, which $\text{PAD}(T, x, v)$ requires to have a timestamp greater than $\text{rver}[T]$. Out of all primitive commands by other threads, we only consider the ones at line 28 and at line 40 (the rest of primitive commands preserve $\text{PAD}(T, x, v)$ trivially).

Firstly, we consider the graph update $\text{NTXWRITE}(n', x)$, which adds a new non-transactional node $n'$ writing to $x$. Since we only consider data-race free histories in our proof of TL2, either $n' \xrightarrow{\text{HB}} T$ or $T \xrightarrow{\text{HB}} n'$ must occur in $G'$. The former cannot be the case, since $n'$ is a new node, and the graph update only adds edges ending in $n'$. Also, $T$ is a live transaction, so by INV.8(d), $T \xrightarrow{\text{HB}} n'$ cannot be in $G'$.

Secondly, we consider a primitive command at line 28, which happens when a transaction $T'$ writes to the register $x$ and results in adding a corresponding write response into the history. Hence, after this command, $\text{writes}(T', x, \_)$ and $\neg \text{aborted}(T')$ both hold. Note that when $T$ reads a non-initial value written by a node $n$, $\neg(T' \xrightarrow{\text{WW}_x} n)$ also holds, since $T'$ is not visible. Additionally, $T'$ is also live, meaning that it has not generated its write timestamp yet, i.e., $\text{wver}[T'] = \top$ holds. Therefore, $\text{PAD}(T, x, v)$ holds too.

Thirdly, we now consider the primitive command at line 40 occurring in some transaction $T'$ writing to $x$. The primitive command replaces previously maximal write timestamp $\text{wver}[T']$ with the incremented value of $\text{clock}$. By INV.7(b), all other read and write timestamps, such as $\text{rver}[T]$, are less or equal to $\text{clock}$ in $(s, H, G)$. Hence, $\text{rver}[T] < \text{wver}[n']$ holds after the primitive command executes.

Lastly, we consider $\text{TXVIS}(n')$ occurring in a transaction $n'$ writing to $x$. Knowing that $(s, H, G) \in \text{PAD}(T, x, v)$, we get that $\text{rver}[T] < \text{wver}[n']$ holds. Since the graph update simply makes $n'$ visible, $\text{PAD}(T, x, v)$ holds. □

We now prove the property of anti-dependencies added at the graph update $\text{TXREAD}(T, x, v)$.

**Proposition C.4.** *If $(s, H, G) \in \text{INV}$, $T \in \text{txns}(H)$ and $(s', H', G')$ is a result of a graph update $\text{TXREAD}(T, x, v)$, then for every anti-dependency $T \xrightarrow{\text{WR}_x} n$ added by the update, both $n \in \text{txns}(H')$ and $\text{rver}[T] < \text{wver}[n]$ hold of $(s', H', G')$.*

*Proof.* Before the graph update, the read operation stores the value of $\text{ver}[x]$ into a local variable $\text{ts1}$, then it stores $\text{reg}[x]$ in a local variable $\text{value}$ as well as the lock status $\text{lock}[x].\text{test}()$ in $\text{locked}$. Afterwards, the read operation stores the value of $\text{ver}[x]$ once again into a local variable $\text{ts2}$. The graph update $\text{TXREAD}(T, x)$ happens provided that the following holds of $(s, H, G)$ prior to the update:

$$\neg \text{locked} \land \text{ts1} = \text{ts2} \land \text{ts2} \leq \text{rver}[T] \quad (6)$$

Note that by checking the above, the read operation ensures that there was a moment in between the two accesses to $\text{ver}[x]$, when $\text{ver}[x] = \text{ts1} = \text{ts2}$, $\text{reg}[x] = \text{value}$ and $\text{lock}[x] = \bot$ all simultaneously held. Let $(s'', H'', G'')$ correspond to that moment. Since that is the past of the execution, $(s'', H'', G'') \in \text{INV}$ holds.

In the following, we demonstrate that $\text{PAD}(T, x, v)$ holds of $(s'', H'', G'')$. Once that is established, Proposition C.3 will ensure that $\text{PAD}(T, x, v)$ is not invalidated till the read operation returns and the graph update executes.

Let us assume that $T$ reads the initial value, i.e. $\text{value} = v_{\text{init}}$. Note that $\text{reg}[x] = v_{\text{init}}$ holds of $(s'', H'', G'')$. By INV.8(a,b), there is no visible node in $G''$ that writes to $x$. Hence, $\text{PAD}(T, x, v)$ holds trivially of $(s'', H'', G'')$. Now let us consider the case when $T$ does not read an initial value, i.e. $\text{value} \neq v_{\text{init}}$. Note that $\text{reg}[x] = \text{value} \neq v_{\text{init}}$ holds of $(s'', H'', G'')$. By INV.8(b), $n$ is last in $\text{WW}_x$. Once again, $\text{PAD}(T, x, v)$ holds trivially of $(s'', H'', G'')$, since no node $n'$ follows $n$ in $\text{WW}_x$. □

**Corollary C.5.** *If $(s, H, G) \in \text{INV}$, $T \in \text{txns}(H)$ and $(s', H', G')$ is a result of a graph update $\text{TXREAD}(T, x, v)$, then for every anti-dependency $T \xrightarrow{\text{WR}_x} n$ added by the update, both $n \in \text{txns}(H')$ and $\text{rver}[T] < \text{wver}[n]$ hold of $(s', H', G')$.*

*Proof sketch.* Note that the graph update $\text{TXREAD}(T, x, v)$ adds the following anti-dependency edges into $G$:

- if $v = v_{\text{init}}$, then $\{T \xrightarrow{\text{RW}_x} n \mid \text{vis}(n) \land \text{writes}(n, x, \_)\}$;
- otherwise, $\{T \xrightarrow{\text{RW}_x} n' \mid \text{writes}(n', x, v) \land n' \xrightarrow{\text{WW}_x} n\}$.

Thus, it is sufficient to prove that $\text{PAD}(T, x, v)$ holds of $(s', H', G')$. This follows from Proposition C.4. □

## C.6 Preservation of INV.2

It is easy to see that the invariants INV.2(b,c) hold by construction of the history. In the following, we argue that INV.2(a) is preserved throughout each execution.

In order to ensure consistency of a history $H$, according to Definition 6.2, we need to demonstrate the following for every read request $\alpha = (\_, \_, \text{read}(x))$ and its matching response $\alpha' = (\_, \_, \text{ret}(v))$:

- when $(\alpha, \alpha') \in \text{Local}(H)$ and performed by a transaction $T$, $v$ is the value written by the most recent write $(\_, \_, \text{write}(x, v))$ preceding the read in $T$;
- when $(\alpha, \alpha') \notin \text{Local}(H)$, there exists a non-local $\beta$ not located in an aborted or live transaction and such that $\beta <_{\text{wr}_x(H)} \alpha'$; if there is no such write, $v = v_{\text{init}}$.

It is easy to see that it is sufficient to ensure preservation of consistency only when we add read-response actions into history, which happens at line 16 or line 24 of the read function.

Let us consider any transaction $T$ that initiated a read operation from a register $x$ (the case of a non-transactional read is analogous to a non-local read by a transaction). The last action by $T$ in history is a read-request $\alpha = (\_, \_, \text{read}(x))$. Let a corresponding read response be $\alpha' = (\_, \_, \text{ret}(v))$ and consider the moment it is added into history $H$, which results in a history $H'$.

We first consider the case when $(\alpha, \alpha') \in \text{Local}(H)$ holds. By Definition 6.1, $T$ writes to $x$ before $\alpha$. Let $v'$ be the value of the last such write; then $\text{writes}(T, x, v')$ holds of $H$ and $H'$. By INV.2(b), $(x, v') \in \text{wset}[T]$. Note that $(x, v') \in \text{wset}[T]$ holds during the entire read operation, because $\text{wset}[T]$ is a local variable of transaction $T$. It is easy to see that in this case the read operation returns a value from the write-set at line 16, meaning that $v = v'$.

We now consider the case when $(\alpha, \alpha') \notin \text{Local}(H)$ holds. By Definition 6.1, $T$ does not write to $x$ before $\alpha$, meaning that there is no value $v'$ such that $\text{writes}(T, x, v')$ holds of history during the execution of the read operation. Hence, by INV.2(b), $\text{wset}[T]$ does not contain a value for $x$. Moreover, the latter is the case during the entire read operation, because $\text{wset}[T]$ is a local variable of transaction $T$. It is easy to see that in this case the read operation at line 24 returns the value $v$ read directly from the register; however, the result of the operation is determined at line 20. Based on INV.8(a) (note that the read operation ensures that $\text{lock}[x]$ is unlocked), the following two observations can be made about the value $v$ at the moment of executing line 20:

1. $v \neq v_{\text{init}}$ if and only if there exists a node $n$ such that $\text{isLastIn}(\text{WW}_x, n)$ and $\text{writes}(n, x, v)$ both hold;
2. $v = v_{\text{init}}$ if and only if there is no visible node writing to $x$.

First, let us consider the case when there exists a node $n$ such that $\text{isLastIn}(\text{WW}_x, n)$ and $\text{writes}(n, x, v)$ both hold. Since it is ordered by $\text{WW}_x$, by Definition 6.3, it is visible, and, therefore, does not denote a live or an aborted transaction. According to $\text{writes}(n, x, v)$, $n$ contains a non-local write action $\beta$ writing $v$ to $x$. Thus, $\beta <_{\text{wr}_x(H')} \alpha'$ holds. Overall, in this case, if $(\alpha, \alpha') \notin \text{Local}(H')$ and $v \neq v_{\text{init}}$, then there exists a non-local $\beta \in n$, which is not an aborted or live transaction, such that $\beta <_{\text{wr}_x(H)} \alpha'$.

We now consider the case when there does not exist a visible node writing to $x$, which is when $v = v_{\text{init}}$. Thus, any node $n$ writing to $x$ must be either a live, aborted or commit-pending non-visible transaction. Additionally, in case of the latter, if $\beta$ is the non-local write to $x$ by $n$, $\beta <_{\text{wr}_x(H')} \alpha'$ does not hold, since $\alpha'$ returns $v = v_{\text{init}}$ and $\beta$ writes a non-initial value[4]. Overall, in this case, if $(\alpha, \alpha') \notin \text{Local}(H')$ and $v = v_{\text{init}}$, then there does not exists a non-local $\beta$ not located in an aborted or live transaction and such that $\beta <_{\text{wr}_x(H)} \alpha'$.

## C.7 Preservation of INV.3

**Proposition C.6.** *If $(s, H, G) \in \text{INV}$ and $(s', H', G')$ is a result adding a new node $n$ by graph updates* TXBEGIN$(n)$, NTXREAD$(n, \_)$ *or* NTXWRITE$(n, \_)$, *the following observation holds of $(s', H', G')$:* $\neg \exists n'. n \xrightarrow{\text{HB} \cup \text{WR} \cup \text{WW} \cup \text{RW}} n'$.

*Proof sketch.* The proof of the proposition is almost trivial, as TXBEGIN$(n)$ and NTXWRITE$(n, \_)$ always order new graph nodes after existing ones. The same is true of NTXREAD$(n, x)$ as well (for every register $x$), however, it is necessary to demonstrate that this graph update does not introduce anti-dependencies originating in $n$.

We consider the register $x$. Let us first assume that it is locked at the moment of the graph update NTXREAD$(n, x)$, i.e., there exists a transaction $T$ such that $\text{lock}[x] = T$. By INV.8(e), $T$ writes to $x$, and, therefore, conflicts with $n$. Since we only consider data-race free histories, it must be the case that either $T \xrightarrow{\text{HB}} n$ or $n \xrightarrow{\text{HB}} T$. However, neither is possible: the former, because according to INV.8(d,e) $T$ is not completed and therefore cannot have outgoing happens-before edges, and the latter, because $n$ is a fresh node, and NTXREAD$(n, x)$ does not introduce edges of the form $n \xrightarrow{\text{HB}} \_$. Thus, we obtained a contradiction. We conclude that $x$ is not locked at the moment of the graph update.

When $\text{lock}[x] = \bot$, by INV.8(a), the value of $\text{reg}[x]$ is either the value written by the last node in $\text{WW}_x$ or $v_{\text{init}}$, if there is no such node. Thus, it is easy to see that in both cases no anti-dependency of the form $n \xrightarrow{\text{RW}_x} \_$ is added into $G$ by NTXREAD$(n, x)$.  □

**Proposition C.7.** *If $(s, H, G) \in \text{INV}$ and $(s', H', G')$ is a result of a graph update, then $(s', H', G')$ satisfies* INV.3.

*Proof.* We consider all possible graph updates separately in the four following cases.

CASE 1: non-transactional graph update NTXWRITE$(v, \_)$ or NTXREAD$(v, \_)$. Both updates add a new node $v$ to the graph, which by Proposition C.6 does not have outgoing edges. Hence, it is easy to see that (HB ; (WR∪WW∪RW)) remains irreflexive after either of the graph updates.

---

[4] See well-formedness conditions for traces in errata of Section 2.2

Case 2: transactional graph update TXBEGIN($T$). This update adds a new node $T$ to the graph and happens-before edges of the following form: _ $\xrightarrow{\text{HB}}$ $T$. Note that by Proposition C.6, no edge starts from $T$. Hence, it is easy to see that (HB ; (WR∪WW∪RW)) remains irreflexive after the graph update.

Case 3: transactional graph update TXVIS($T$). It adds edges only of the following form: _ $\xrightarrow{\text{WW}}$ $T$ and _ $\xrightarrow{\text{RW}}$ $T$. Adding such edges cannot invalidate irreflexivity of (HB ; (WR ∪ WW ∪ RW)), unless there is a node $n$ such that $T \xrightarrow{\text{HB}} n$ holds of $(s, H, G)$. However, by invariant INV.8(d), HB-edges do not originate in $T$, so we can conclude that (HB ; (WR∪WW∪RW)) remains irreflexive after TXVIS($T$).

Case 4: transactional graph update TXREAD($T, x, \_$). According to Figure 10, it adds edges of the form _ $\xrightarrow{\text{WR}_x}$ $T$, $T \xrightarrow{\text{RW}_x}$ _ and _ $\xrightarrow{\text{HB}} T$. Note that the transaction $T$ is not completed at the moment $(s, H, G)$. Hence, by invariant INV.8(d), HB-edges do not originate from $T$ prior to the update. The transaction $T$ is also not visible at the moment of the graph update, so WW and WR-edges do not originate from $T$ either according to Definition 6.3. Therefore, the only kind of edges going from $T$ in $G$ and $G'$ is anti-dependencies of the form $T \xrightarrow{\text{RW}}$ _. We need to demonstrate that there is no node $n$ such that a cycle $T \xrightarrow{\text{RW}} n \xrightarrow{\text{HB}} T$ appears in $G'$ after the graph update. To this end, we consider three possibilities:

1. only the edge $T \xrightarrow{\text{RW}_x} n$ is added by the graph update, and $n \xrightarrow{\text{HB}} T$ is present in $G$;
2. only the edge $n \xrightarrow{\text{HB}} T$ is added by the graph update, and $T \xrightarrow{\text{RW}_y} n$ is present in $G$ (for some register $y$);
3. edges $T \xrightarrow{\text{RW}_x} n$ and $n \xrightarrow{\text{HB}} T$ are both added by the graph update.

We start with the first potential cycle, and demonstrate by contradiction that it never takes place. By Proposition C.4, $n$ is a transaction and rver[$T$] < wver[$n$]. Since $n$ is a transaction, Lemma B.11 gives us that there is a path $n \xrightarrow{\text{RT∪TXWR}}^+ T$ in the graph $G$. Recall that $T$ is not visible and, according to the definition of TXREAD($T, x, \_$) in Figure 10, $n$ is visible. By applying Proposition C.8 to the path $n \xrightarrow{\text{RT∪TXWR}}^+ T$, we learn that either wver[$n$] ≤ rver[$T$] or rver[$T$] = ⊥ holds of $(s, H, G)$. However, at the moment of the graph update, transactions already have generated their read timestamps, meaning that only wver[$n$] ≤ rver[$T$] can be the case. Thus, we arrived to a contradiction.

We now consider the second potential cycle, and demonstrate by contradiction that it is never added. Let us assume that $T \xrightarrow{\text{RW}_y} n$ is in the graph $G$. Hence, by definition of RW$_y$, $T$ must have already read from $y$, from which we conclude that $y \neq x$. We also assume that the edge $n \xrightarrow{\text{hb}} T$ is added by the graph update, which, according to Figure 10, only happens when $T'$ is a transaction from which $T$ reads from and there is a node $n'$ in the thread of $T'$ such that the following configuration takes place:

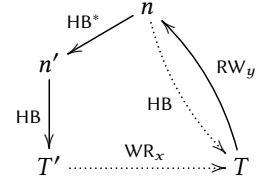

where the solid lines depict edges present in the original graph $G$, and the dotted lines are new edges added by the graph update. We consider two possibilities depending on whether $n$ represents a transaction or not. Let us first assume it is a transaction. By Lemma B.11, $n \xrightarrow{\text{RT∪TXWR}}^+ T'$ holds of the graph $G$. By applying Proposition C.8 to the path $T \xrightarrow{\text{RW}_y} n \xrightarrow{\text{RT∪TXWR}}^+ T'$, we learn that rver[$T$] < wver[$T'$]. However, the graph update happens only if wver[$T'$] ≤ rver[$T$], meaning that this possibility never arises. Let us now assume that $n$ is non-transactional. Since $H$ is data-race free, either $T \xrightarrow{\text{HB}} n$ or $n \xrightarrow{\text{HB}} T$ is the case for the graph $G$. Knowing that $(s, H, G) \in$ INV.3 and that $T \xrightarrow{\text{RW}_y} n$ is in $G$, we conclude that only $T \xrightarrow{\text{HB}} n$ is possible. Hence, $T \xrightarrow{\text{HB}} T'$ holds, and, by Lemma B.11, so does $T \xrightarrow{\text{RT∪TXWR}}^+ T'$. By applying Proposition C.8 to the latter path, we learn that rver[$T$] < wver[$T'$]. However, the graph update happens only if wver[$T'$] ≤ rver[$T$], meaning that this possibility never arises.

Finally, we consider the last possible cycle, and demonstrate by contradiction that it is never added. Let us assume that the edges $T \xrightarrow{\text{RW}_x} n$ and $n \xrightarrow{\text{HB}} T$ are both added by the graph update and form a cycle. This is only possible if the following configuration (analogous to the case of the second cycle) takes place:

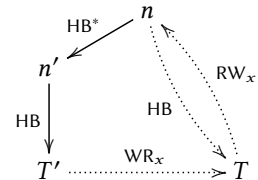

where the solid lines depict edges present in the original graph $G$, and the dotted lines are new edges added by the graph update. By Proposition C.4, $n$ is a transaction and rver[$T$] < wver[$n$]. By Lemma B.11, $n \xrightarrow{\text{RT∪TXWR}}^+ T'$ holds of the graph $G$. Note that $n$ is visible transaction, and $T$ is not. By applying Proposition C.8 to the path $n \xrightarrow{\text{RT} \cup \text{TXWR}}{}^+ T' \xrightarrow{\text{WR}_x} T$, we learn that either $\text{wver}[n] \leq \text{rver}[T]$ or $\text{rver}[T] = \bot$. However, the latter cannot be the case at the moment of the graph update. Hence, we obtained a contradiction: $\text{rver}[T] < \text{wver}[n]$ and $\text{wver}[n] \leq \text{rver}[T]$ both hold. □

### C.8 Preservation of INV.5

**Proposition C.8.** *If $(s, H, G) \in \text{INV}$, $T \in \text{txns}(H)$ and $(s', H', G')$ is a result of a graph update, then $(s', H', G')$ satisfies INV.5.*

*Proof.* Graph updates NTXREAD(_, _) and NTXWRITE(_, _) do not affect the invariant. Therefore, we consider the following cases in the proof: graph updates TXBEGIN($T'$), TXREAD($T', x$), TXVIS($T'$) and also primitive commands at at line 11, line 45 and line 40 in Figure 9. We consider primitive commands in the proof, because the ones at line 11 and line 40 may change the implication of INV.5 for existing nodes and edges satisfying its premise, meaning that we need to prove that the implication remains true. The rest of the graph updates may add new nodes and edges satisfying the premise of the invariant, so we demonstrate that the implication holds of them too.

We first consider a primitive command at line 11 assigning a read version to a transaction $T'$, which happens only when $\text{rver}[T'] = \bot$. As follows from invariants INV.7(c,d,e), in order to have the minimal read timestamp $T'$ must be neither visible nor have non-locally read from any register. This is not the case for transactions in the invariants INV.5(b-e), so we focus on the invariant INV.5(a). Let us consider a transaction $T$ such that $T \xrightarrow{\text{RT}} T'$ holds. By invariant INV.7(b), $\text{rver}[T] \leq \text{clock}$ and either $\text{wver}[T] = \top$ or $\text{wver}[T] \leq \text{clock}$. Let us assume that $\text{vis}(T)$ holds. By INV.7(e), $\text{wver}[T] \neq \top$, so only $\text{wver}[T] \leq \text{clock}$ is possible. Since the new value of $\text{rver}[T']$ is clock, it is the case that $\text{wver}[T] \leq \text{rver}[T']$, meaning that the invariant INV.5(a) is preserved. Let us now consider the case when $\text{vis}(T)$ does not hold. Knowing that $\text{rver}[T] \leq \text{clock}$ and that the new value of $\text{rver}[T']$ is clock, we conclude $\text{rver}[T] \leq \text{rver}[T']$, which means that the invariant INV.5(a) is preserved.

Second, we consider a primitive command at line 40 assigning a write timestamp to a transaction, which happens only when the transaction has the write timestamp $\top$. The contra-positive of INV.7(e) states that such transaction is not visible. Note that every occurrence of $\text{wver}[\cdot]$ in INV.5 corresponds to a visible transaction. Therefore, line 40 does not invalidate INV.5.

Third, we consider a graph update TXBEGIN($T'$). This graph update adds a new node corresponding to the transaction $T'$. We also need to consider real-time order between completed transactions and $T'$. Let $T$ be any completed transaction such that $T \xrightarrow{\text{RT}} T'$. Since $\text{rver}[T']$ is initialized with $\text{rver}[T'] = \bot$, INV.6(a) holds trivially.

Forth, we consider a graph update TXREAD($T', x, \_$) adding read and anti-dependencies between transaction. Let us first consider the new read dependency $T \xrightarrow{\text{WR}_x} T'$. Since the graph update only happens if $\text{wver}[T] \leq \text{rver}[T']$ holds, the invariant INV.5 is not invalidated by adding the new read dependency. We now consider any new anti-dependency $T' \xrightarrow{\text{RW}_x} T''$. By Proposition C.4, $\text{rver}[T'] < \text{wver}[T'']$, meaning that INV.5(d) is preserved too.

Fifth, we consider a graph update TXVIS($T'$). Let us consider any new edge $T \xrightarrow{\text{WW}_x \cup \text{RW}_x} T'$. Since the transaction $T'$ first locks registers of its write-set and then performs TXVIS($T'$), $\text{lock}[x] = T'$ holds at the moment of the graph update. Preservation of INV.5(c–e) by the graph update then follows immediately from INV.6(a–c) accordingly.

Finally, we consider a possibility of a primitive command by $T$ at line 45 setting $\text{pv}[T][x]$ to true for some register $x$. Let us assume that there exists a transaction $T'$ such that $T \xrightarrow{\text{RW}_x} T'$ holds prior to the execution of line 45 (if there are several such transactions, we let $T'$ be the one occurring the last in $\text{WW}_x$). By INV.5(d), $\text{rver}[T] < \text{wver}[T']$ holds then. Note that line 45 sets $\text{pv}[T][x]$ to true only if $\text{lock}[x] = \bot$ and $\text{ts} \leq \text{rver}[T]$, where $\text{ts} = \text{ver}[x]$. By INV.8(b), $\text{ver}[x] = \text{wver}[T']$. As a consequence, $\text{wver}[T'] \leq \text{rver}[T]$ holds, which contradicts our previous observation about $\text{wver}[T']$ and $\text{rver}[T]$. Hence, there does not exist a transaction $T'$ such that $T \xrightarrow{\text{RW}_x} T'$, and INV.5(e) is preserved by the primitive command at line 45. □

### C.9 Preservation of INV.4

In this section, we demonstrate that all graph updates preserve the invariant INV.4. To this end, firstly, we observe that graph updates NTXREAD(_, _) and NTXWRITE(_, _) do not introduce edges between pairs of transactions and therefore cannot possibly invalidate INV.4. Secondly, we argue that the graph update TXINIT($T$), which adds a new transaction $T$ into the graph and implies new real-time order edges between transactions ending in $T$, also straightforwardly preserves INV.4. As we previously demonstrated in Proposition C.6, no edges originate from the new transaction $T$, so it is easy to see that the graph update does not introduce any cycle and, thus, preserves INV.4. In the rest of the section, we consider the remaining graph updates TXREAD(_, _) and TXVIS(_), we prove Proposition C.9 and Proposition C.10.

**Proposition C.9.** *If $(s, H, G) \in \text{INV}$, $T' \in \text{txns}(H)$ and $(s', H', G')$ is a result of a graph update TXREAD($T', x$), then $(s', H', G')$ satisfies INV.4.*

*Proof.* Before the graph update, the read operation stores the value of $\text{ver}[x]$ into a local variable ts1, then it stores

reg$[x]$ in a local variable value as well as the lock status lock$[x]$.test() in locked. Afterwards, when the read operation stores the value of ver$[x]$ once again into a local variable ts2, the graph update TXREAD$(T', x)$ may happen, provided that the following holds of $(s, H, G)$ prior to the update:

$$\neg\text{locked} \wedge \text{ts1} = \text{ts2} \wedge \text{ts2} \leq \text{rver}[T'] \qquad (7)$$

Note that by checking the above, the read operation ensures that there was a moment in between the two accesses to ver$[x]$, when ver$[x]$ = ts1 = ts2, reg$[x]$ = value and lock$[x]$ = $\bot$ all simultaneously held. Let $(s'', H'', G'')$ correspond to that moment.

Let us assume that the condition (7) holds, and the graph update takes place. Since TXREAD$(T', x)$ adds different edges depending on whether value = $v_{\text{init}}$ holds, we consider these two cases in the proof separately.

Firstly, we consider the case when value = $v_{\text{init}}$ holds. According to Figure 10, only RW edges are added into the graph in this case. We prove that such edges do not create cycles by contradiction. Let us imagine a cycle in the graph $G'$ going through a new anti-dependency edge $T' \xrightarrow{\text{RW}_x} T''$. The path $T'' \xrightarrow{\text{RT} \cup \text{WR} \cup \text{WW} \cup \text{RW}}{}^+ T'$ is present in the graph $G$. Hence, after applying Proposition C.2 to visible $T''$ and non-visible $T'$, we obtain that wver$[T''] \leq$ rver$[T']$. Additionally, by Proposition C.4, the anti-dependency $T' \xrightarrow{\text{RW}_x} T''$ implies that rver$[T']$ < wver$[T'']$. Overall, we obtained a contradiction.

Secondly, we consider the case when value $\neq v_{\text{init}}$. According to Figure 10, HB, RW and WR edges are added into the graph. However, in this proof we do not consider the new HB edges, since they cannot cause cycles invalidating INV.4. For the same reason, we only consider WR-edges originating from transactions. Let $T \xrightarrow{\text{WR}_x} T'$ be the read dependency added and let $T' \xrightarrow{\text{RW}_x} T''$ be any of the new anti-dependencies added by the graph update. We prove by contradiction that neither of them causes a cycle in $(\text{RT} \cup \text{txWR} \cup \text{txWW} \cup \text{txRW})$.

The invariant INV.8(a) asserts that the value of the register reg$[x]$ previously stored in a local variable value is the value written by the last transaction in WW$_x$. From this observation, it is easy to conclude that $T$ is this transaction at the moment $(s'', H'', G'')$. By INV.8(b), ver$[x]$ coincides with the write version wver$[T]$. Knowing that the condition (7) holds, we conclude that so does wver$[T] \leq$ rver$[T']$. Also, by Proposition C.4, the anti-dependency $T' \xrightarrow{\text{RW}_x} T''$ implies that rver$[T']$ < wver$[T'']$.

Adding edges $T \xrightarrow{\text{WR}_x} T'$ or $T' \xrightarrow{\text{RW}_x} T''$ may cause three kinds of cycles: the ones containing exactly one of the two edges, and the one containing both. We consider these cases separately:

**Case #1:** $T \xrightarrow{\text{WR}_x} T'$. Previously, we have shown that wver$[T] \leq$ rver$[T']$ holds. Also, by INV.7(d), rver$[T'] \neq \bot$ holds. Overall, both $\neg(\text{rver}[T'] <$ wver$[T])$ and rver$[T'] \neq \bot$ hold. Note that $T'$ is not visible and $T$ is. When all of the aforementioned takes place, the contra-positive of Proposition C.2 states that there is no path $T' \xrightarrow{\text{RT} \cup \text{WR} \cup \text{WW} \cup \text{RW}}{}^+ T$. Hence, the edge $T \xrightarrow{\text{WR}_x} T'$ alone does not cause a cycle in $\text{RT} \cup \text{txWR} \cup \text{txWW} \cup \text{txRW}$.

**Case #2:** $T' \xrightarrow{\text{RW}_x} T''$. Previously, we have shown that rver$[T']$ < wver$[T'']$ holds. Also, by INV.7(d), rver$[T'] \neq \bot$ holds. Overall, both $\neg(\text{wver}[T''] \leq$ rver$[T'])$ and rver$[T'] \neq \bot$ hold. Note that $T''$ is visible and $T'$ is not. When all of the aforementioned takes place, the contra-positive of Proposition C.2 states that there is no path $T'' \xrightarrow{\text{RT} \cup \text{WR} \cup \text{WW} \cup \text{RW}}{}^+ T'$. Hence, the edge $T' \xrightarrow{\text{RW}_x} T''$ alone does not cause a cycle in $\text{RT} \cup \text{txWR} \cup \text{txWW} \cup \text{txRW}$.

**Case #3:** $T \xrightarrow{\text{WR}_x} T' \xrightarrow{\text{RW}_x} T''$. Previously, we have shown that wver$[T] \leq$ rver$[T']$ < wver$[T'']$ holds. Note that $T''$ and $T$ are both visible. As the contra-positive of Proposition C.2 states, if wver$[T''] <$ wver$[T]$ does not hold, then there is no path $T'' \xrightarrow{\text{RT} \cup \text{WR} \cup \text{WW} \cup \text{RW}}{}^+ T$. Hence, the edges $T \xrightarrow{\text{WR}_x} T' \xrightarrow{\text{RW}_x} T''$ do not cause a cycle in $\text{RT} \cup \text{txWR} \cup \text{txWW} \cup \text{txRW}$.

□

**Proposition C.10.** *If* $(s, H, G) \in \text{INV}$, $T \in \text{txns}(H)$ *and* $(s', H', G')$ *is a result of a graph update* TXVIS$(T')$, *then* $(s', H', G')$ *satisfies* INV.4.

*Proof.* Note the graph update TXVIS$(T')$ occurs after line 50 in the transaction $T'$, meaning that the following holds of $T'$ in $(s, H, G)$:

- $\forall(x, \_) \in \text{wset}[T']. \text{lock}[x] = T'$;
- $\forall(x, \_) \in \text{rset}[T']. \text{pv}[T'][x] = \text{true}$.

The graph update TXVIS$(T')$ adds the following new edges into the opacity graph $G = (N, \text{vis}, \text{HB}, \text{WR}, \text{WW}, \text{RW})$ for every register $x$ such that $(x, \_) \in \text{wset}[T']$ holds:

- $\{n \xrightarrow{\text{WW}_x} T' \mid n \in N \wedge n \neq T' \wedge \text{vis}(n) \wedge \text{writes}(n, x, \_)\}$
- $\{n \xrightarrow{\text{RW}_x} T' \mid n \in N \wedge n \neq T' \wedge \text{reads}(n, x, \_)\}$

The update also makes $T'$ visible in $(s', H', G')$. It is easy to see that only new edges between transactions can invalidate INV.4, since it asserts absence of cycles over transactions only. As the new dependencies end in the same node $T'$, they will not appear in the same simple cycle. These two observations together allow us to consider each edge individually and prove that it does not create a cycle in $(\text{RT} \cup \text{txWR} \cup \text{txWW} \cup \text{txRW})$.

**WW-edges.** We show that adding an edge $T \xrightarrow{WW_x} T'$ into the opacity graph $G$ preserves INV.4.

Note that $\text{lock}[x] = T'$ and that $T$ is such that $\text{vis}(T)$, $T \neq T'$ and $\text{writes}(T, x, \_)$ all hold of $(s, H, G)$; then from INV.6(a) we infer that $\text{wver}[T] < \text{wver}[T']$. The latter also holds of $(s', H', G')$, as the graph update does not change write timestamps of transactions.

We demonstrated in Proposition C.8 that $(s', H', G')$ satisfies INV.5 and argued that it also satisfies INV.8. This enables applying Proposition C.2 to $(s', H', G')$. Note that $\text{vis}(T)$ and $\text{vis}(T')$ both hold of $(s', H', G')$. The contra-positive of Proposition C.2 asserts that $\text{vis}(T)$, $\text{vis}(T')$ and $\text{wver}[T] < \text{wver}[T']$ together imply that $T' \xrightarrow{RT \cup WR \cup WW \cup RW}{}^+ T$ does not hold, meaning that adding the edge $T \xrightarrow{WW_x} T'$ into $G$ does not create cycles in $RT \cup \text{txWR} \cup \text{txWW} \cup \text{txRW}$.

**RW-edges.** We show that adding the edge $T \xrightarrow{RW_x} T'$ into the opacity graph $G$ preserves INV.4. We consider the triple $(s, H, G)$ of the state prior to the graph update and split the proof in two cases depending on whether $T$ is visible or not.

First, we consider the case when $T$ is visible. By INV.8(c), $\text{pv}[T][x] = \text{true}$ holds of $(s, H, G)$. Also, recall that $T'$ holds a lock on $x$. When that is the case, according to INV.6(c), $\text{wver}[T] < \text{wver}[T']$ holds. The latter also holds of $(s', H', G')$, as the graph update does not change write timestamps of transactions.

We demonstrated in Proposition C.8 that $(s', H', G')$ satisfies INV.5 and argued that it also satisfies INV.8. This enables applying Proposition C.2 to $(s', H', G')$. Note that $\text{vis}(T)$ and $\text{vis}(T')$ both hold of $(s', H', G')$. The contra-positive of Proposition C.2 asserts that $\text{vis}(T)$, $\text{vis}(T')$ and $\text{wver}[T] < \text{wver}[T']$ together imply that $T' \xrightarrow{RT \cup WR \cup WW \cup RW}{}^+ T$ does not hold, meaning that adding the edge $T \xrightarrow{RW_x} T'$ into $G$ does not create cycles in $RT \cup \text{txWR} \cup \text{txWW} \cup \text{txRW}$.

We now return to the case when $T$ is not visible. Recall that $T'$ holds a lock on $x$. When that is the case, according to INV.6(b), $\text{rver}[T] < \text{wver}[T']$ holds of $(s, H, G)$. The latter also holds of $(s', H', G')$, as the graph update does not change write timestamps of transactions.

We demonstrated in Proposition C.8 that $(s', H', G')$ satisfies INV.5 and argued that it also satisfies INV.8. This enables applying Proposition C.2 to $(s', H', G')$. Note that $\neg\text{vis}(T)$ and $\text{vis}(T')$ both hold of $(s', H', G')$. Also, $\text{rver}[T] \neq \bot$, since by INV.7(d) transaction satisfying $\text{reads}(T, x, \_)$ already has its read timestamp initialized. The contra-positive of Proposition C.2 asserts that $\text{vis}(T')$, $\neg\text{vis}(T)$ and $\neg(\text{wver}[T'] \leq \text{wver}[T] \lor \text{rver}[T] = \bot)$ together imply that $T' \xrightarrow{RT \cup WR \cup WW \cup RW}{}^+ T$ does not hold, meaning that adding the edge $T \xrightarrow{RW_x} T'$ into $G$ does not create cycles in $RT \cup \text{txWR} \cup \text{txWW} \cup \text{txRW}$.  □

### C.10  Preservation of INV.6

**Proposition C.11.** *When $(s, H, G) \in \text{INV}$, $T \in \text{txns}(H)$ and $(s', H', G')$ is a result of execution of line 45 by $T$, if $\text{lock}[x]$ is held by some transaction $T'$, then $\text{rver}[T] < \text{wver}[T']$.*

*Proof sketch.* The transaction $T$ post-validates its read from $x$ at line 45 simultaneously with loading the value of $\text{ver}[x]$. At the previous line, the commit operation of $T$ stores the value of $\text{lock}[x]$ in a local variables $\text{locked}$. The post-validation is successful only if $\neg\text{locked}$ and $\text{ver}[x] \leq \text{rver}[T]$. Note that this check has already been performed when $T$ performed a read operation from $x$. This way TL2 ensures that there has been a moment between the execution of line 20 in the read operation and line 45, at which both $\text{lock}[x] = \bot$ and $\text{ver}[x] \leq \text{rver}[T]$ hold simultaneously. Let $(s'', H'', G'')$ correspond to that moment.

For convenience, we introduce a predicate $\text{PVP}(T, x)$, which holds of $(s, H, G)$ whenever the following does: $\text{reads}(T, x, \_)$ holds, and if $\text{lock}[x]$ is held by some transaction $T'$, then $\text{rver}[T] < \text{wver}[T']$. It is easy to see that $\text{PVP}(T, x)$ holds trivially of $(s'', H'', G'')$, since $\text{lock}[x]$ is not held at that moment. We further demonstrate that $\text{PVP}(T, x)$ is never invalidated by transitions and graph updates of $T$ or by other threads. This allows us to conclude that $\text{PVP}(T, x)$ holds of $(s', H', G')$.

Note that once $\text{PVP}(T, x)$ is established, it could only be possibly invalidated by:

- a transition by $T'$ at line 33 acquiring the lock $\text{lock}[x]$;
- a transition by $T'$ at line 40 setting the write timestamps $\text{wver}[T']$ from $\top$ to the incremented clock value.

However, both preserve $\text{PVP}(T, x)$. Indeed, when the former transition occurs, $\text{wver}[T'] = \top$ holds. Since $\top$ is the maximal possible timestamp, $\text{rver}[T] < \text{wver}[T']$ holds then. When the second transition occurs in a transaction $T'$ holding a lock on $x$, by INV.7(b), we observe that $\text{rver}[T] < \text{clock} + 1$ and that $\text{wver}[T'] = \text{clock} + 1$, which also allows to conclude $\text{rver}[T] < \text{wver}[T']$.  □

**Proposition C.12.** *If $(s, H, G) \in \text{INV}$, $T \in \text{txns}(H)$ and $(s', H', G')$ is a result of a graph update, then $(s', H', G')$ satisfies INV.6.*

*Proof.* Graph updates $\text{NTXREAD}(\_, \_)$ and $\text{NTXWRITE}(\_, \_)$ do not affect the invariant, since it only asserts properties of transactions. Thus, in this proof we consider the following graph updates and transitions:

1. the graph update $\text{TXVIS}(T)$, which may enable the premise of INV.6(a);
2. the graph update $\text{TXREAD}(T, x, \_)$, which may enable the premise of INV.6(b);
3. the transition by $T$ at line 45 setting $\text{pv}[T][x]$ to $\text{true}$, which may enable the premise of INV.6(c);
4. the transition by $T'$ at line 33 acquiring the lock $\text{lock}[x]$, which may enable the premise of INV.6(a,b,c);

5. the transition by $T$ at line 11 setting the read timestamp rver$[T]$ from $\bot$ to the current clock value, which may affect the implication of INV.6(b);
6. the transition by $T$ (or $T'$) at line 40 setting the write timestamps wver$[T]$ (or wver$[T']$) from $\top$ to the incremented clock value, which may affect the implication of INV.6(a,b,c).

Firstly, we discuss a possibility when the graph update TXVIS($T$) in a transaction $T$ writing to $x$ makes $T$ visible, when lock$[x] = T'$ holds already. Note that TXVIS($T$) only happens when $T$ holds the lock on lock$[x]$, meaning that the possibility in discussion never arises.

Second, we consider a possibility when the graph update TXREAD($T, x, v$) (or rather, the read response added into the history) in a transaction $T$ makes reads($T, x, v$) hold, when lock$[x] = T'$ holds already. In the proof of Proposition C.4, we shown PAD($T, x, v$), meaning that the following holds of $T$ in this case:

- if $T$ reads the initial value, then:

$$\forall n'.\ \text{writes}(n', x, \_) \land \neg \text{aborted}(n') \implies$$
$$(n' \in \text{txns}(H)) \land \text{rver}[T] < \text{wver}[n']$$

- if $T$ reads the value written by a node $n$, then:

$$\forall n'.\ \text{writes}(n', x, \_) \land \neg \text{aborted}(n') \land \neg n' \xrightarrow{\text{WW}_x} n \implies$$
$$(n' \in \text{txns}(H)) \land \text{rver}[T] < \text{wver}[n']$$

We further argue that premises of the two properties above hold of $T'$. Note that $T'$ holds a lock on $x$. By INV.8(e), $T'$ is not completed (and, therefore, not aborted), writes to $x$ and is not followed in $\text{WW}_x$ by any other transaction. Let us show by contradiction that it is also not followed in $\text{WW}_x$ by any non-transactional node either. Let us assume that there is a non-transactional $n$ such that $T' \xrightarrow{\text{WW}_x} n$ holds. By INV.1, the history is DRF, meaning that either $T' \xrightarrow{\text{HB}} n$ or $n \xrightarrow{\text{HB}} T'$ must hold. However, the former contradicts INV.8(d) and the latter contradicts INV.3. Hence, such node $n$ does not exist. Overall, we demonstrated that writes($T', x, \_$), $\neg$aborted($T'$) and $\forall n.\ \neg T' \xrightarrow{\text{WW}_x} n$ all hold, hence, from PAD($T, x, v$) we obtain rver$[T] <$ wver$[T']$.

Third, we consider the transition by $T$ at line 45 setting pv$[T][x]$ to true, when lock$[x] = T'$ holds already. By Proposition C.11, rver$[T] <$ wver$[T']$ holds in this case.

Forth, we consider a possibility of the transition by $T'$ at line 33 acquiring lock$[x]$ and enabling the premise of either of the invariants, which we consider separately. To this end, we start with considering the case when writes($T, x$) and vis($T$) hold, and $T'$ acquires lock$[x]$. We need to demonstrate that wver$[T] <$ wver$[T']$. By INV.7(e), wver$[T] \neq \top$. However, at line 33, wver$[T'] = \top$. Hence, wver$[T] <$ wver$[T']$, which concludes INV.6(a). Let us now consider the case when reads($T, x, \_$) holds, and $T'$ acquires lock$[x]$. We need to demonstrate that rver$[T] <$ wver$[T']$. By INV.7(b), rver$[T]$ is smaller than the value of the global clock, so it cannot be $\top$. Knowing that wver$[T'] = \top$ holds, we get that so does rver$[T] <$ wver$[T']$, which concludes INV.6(b). We now consider the case when pv$[T][x]$ holds, and $T'$ acquires lock$[x]$. We need to demonstrate that wver$[T] <$ wver$[T']$. By INV.7(e), if $T$ post-validated at least one of its reads, wver$[T] \neq \top$. Knowing that at line 33, wver$[T'] = \top$ holds, we get that so does wver$[T] <$ wver$[T']$, which concludes INV.6(c).

Fifth, we consider a possibility of the transition by $T$ at line 11 affecting the implication of INV.6(b). We assume that reads($T, x, \_$) and lock$[x] = T'$ both hold prior to the transition. By INV.7(d), rver$[T] \neq \bot$ then. Since the transition by $T$ at line 11 only happens when rver$[T] = \bot$, it cannot possibly invalidate INV.6(b).

Finally, we consider a possibility of the transition by $T$ or $T'$ at line 40 affecting the implication of INV.6(a, b, c). Let us first show that such transition in fact cannot happen in $T$. For each of the invariants, we assume that its premise holds, and that wver$[T] <$ wver$[T']$ prior to the transition. This means that wver$[T] \neq \top$, since $\top$ is the maximal timestamp value. Knowing that line 40 executes only when wver$[T] = \top$, we conclude that it cannot invalidate INV.6. We now consider line 40 changing the write timestamp wver$[T']$ from $\top$ to (clock+1). For each of the invariants, we assume that its premise holds. Let us show that wver$[T] <$ wver$[T']$ is preserved by the transition. As we have shown, when the aforementioned inequality holds before the transition, wver$[T] \neq \top$. By INV.7(b), the value (clock+1) is greater than every write timestamp distinct from $\top$, such as wver$[T]$. Hence, wver$[T] <$ wver$[T']$ holds after the transition. We now show that rver$[T] <$ wver$[T']$ is preserved by the transition. By INV.7(b), the value (clock+1) is greater than every read timestamp, such as rver$[T]$. Hence, rver$[T] <$ wver$[T']$ holds after the transition. □